\documentclass[preprint,sort&compress,11pt]{elsarticle}

\usepackage{graphicx}
\usepackage{amsmath}

\usepackage{stmaryrd}
\usepackage{fullpage}

\usepackage{subcaption}
\usepackage{enumerate}

\usepackage{amssymb}
\usepackage{amsthm}
\usepackage{bbm}
\usepackage{bm}
\usepackage{url}
\usepackage{listings}
\usepackage{gensymb}
\usepackage{array}

\usepackage{multirow}
\newcolumntype{P}[1]{>{\centering\arraybackslash}m{#1}}

\newcolumntype{L}[1]{>{\raggedright\let\newline\\\arraybackslash\hspace{0pt}}m{#1}}
\newcolumntype{C}[1]{>{\centering\let\newline\\\arraybackslash\hspace{0pt}}m{#1}}
\newcolumntype{R}[1]{>{\raggedleft\let\newline\\\arraybackslash\hspace{0pt}}m{#1}}

\journal{Communications in Computational Physics}

\begin{document}

\begin{frontmatter}
\title{Enforcing imprecise constraints on generative adversarial networks for emulating physical systems}

\author[vt]{Yang Zeng}
\author[vt]{Jin-Long Wu\fnref{fnjl}}
\fntext[fnjl]{Current affiliation: California Institute of Technology, Pasadena, CA 91125, USA.}
\author[vt]{Heng Xiao\corref{mycor}}

\cortext[mycor]{Corresponding author}
\ead{hengxiao@vt.edu}
\address[vt]{Kevin T. Crofton Department of Aerospace and Ocean Engineering, Virginia Tech, Blacksburg, VA 24060, USA}


\begin{abstract}
Generative adversarial networks (GANs) were initially proposed to generate images by learning from a large number of samples.
Recently, GANs have been used to emulate complex physical systems such as turbulent flows. However, a critical question must be answered before GANs can be considered trusted emulators for physical systems: do GANs-generated samples conform to the various physical constraints? These include both deterministic constraints (e.g., conservation laws) and statistical constraints (e.g., energy spectrum of turbulent flows). The latter have been studied in a companion paper (Wu et al., Enforcing statistical constraints in generative adversarial networks for modeling chaotic dynamical systems. Journal of Computational Physics. 406, 109209, 2020). In the present work, we enforce deterministic yet imprecise constraints on GANs by incorporating them into the loss function of the generator. We evaluate the performance of physics-constrained GANs on two representative tasks with geometrical constraints (generating points on circles) and differential constraints (generating divergence-free flow velocity fields), respectively. In both cases, the constrained GANs produced samples that conform to the underlying constraints rather accurately, even though the constraints are only enforced up to a specified interval. More importantly, the imposed constraints significantly accelerate the convergence and improve the robustness in the training, indicating that they serve as a physics-based regularization. These improvements are noteworthy, as the convergence and robustness are two well-known obstacles in the training of GANs. 
\end{abstract}


\begin{keyword}
Generative adversarial networks, physics constraints, physics-informed machine learning
\end{keyword}
\end{frontmatter}


\section{Introduction}
\label{sec:intro}

\subsection{Physical applications of GANs: progress and challenges}
Machine learning and particularly deep learning has achieved significant success in a wide range of commercial domain applications such as image recognition, audio recognition, and natural language processing~\cite{nasrabadi2007pattern,krizhevsky2012imagenet,lecun2015deep, goodfellow2016deep,neal2012bayesian}.  In recent years, machine learning has been widely adopted in scientific applications, leading to an emerging field referred to as \emph{scientific machine learning}. Example scientific applications of machine learning include augmenting or constructing data-driven turbulence models~\cite{wang2017physics,wu2018physics,ling2016reynolds}, generating realistic animations of flows~\cite{eckert2018coupled,ren2018visual,wiewel2020latent,kim2019deep}  discovering or solving differential equations~\cite{brunton2016discovering,long2019pde,sirignano2018dgm,he2019mgnet,berg2018unified,tripathy2018deep,chen2020comparison}.

Recently, generative adversarial networks (GANs)~\cite{goodfellow2014generative} emerged as a promising model in machine learning. GANs construct mappings from a generic (e.g., uniform or Gaussian) probability distribution to the data distribution. Once trained, such models can generate new samples that are not in the training database but conform to the data distribution.
As the training only uses unlabeled data, generative models belong to unsupervised learning.
GANs have shown promises in many scientific applications from synthesizing CT-scan images of rocks~\cite{mosser2017reconstruction,mosser2018stochastic} to generating flow fields~\cite{king2017creating,xie2018tempogan,kim2019deep} or solutions of ordinary, partial or stochastic differential equations~\cite{stinis2019enforcing,farimani2017deep,yang2018physics}.
The successful applications to physics prompt a critical question on the capability of GANs: do they generate samples that conform to the underlying physical constraints? These constraints are implicitly embedded in the training data to certain accuracy, because the data are obtained either by solving the equations that \emph{reflect} these constraints or by directly observing the physical system that \emph{obey} such constraints. However, the constraints are not explicitly encoded in the GANs.
In the above-mentioned applications, the generated output are physical fields that often reside in high-dimensional spaces. This is in contrast to supervised learning, where the output are multi-class labels or low-dimensional scalar and vectors (e.g., the permeability or velocity at a point). Taking the GANs based PDE-emulator for example~\cite{farimani2017deep}, the output temperature field discretized on a mesh of $100 \times 100$ grid points has a dimension of $10^4$. Nevertheless, the physical laws expressed in the form of PDEs place heavy constraints on the admissible solutions. For example, velocity fields of fluid flows must be divergence-free due to mass conservation; temperature fields are typically smooth due to the Laplace operator in the governing equation. Such constraints dictated that admissible (i.e., physical and realistic) solutions must lie on a low-dimensional manifold embedded in a high-dimensional space. 
Therefore, it is imperative to ensure the constraint-respecting properties of GANs before using them as trusted emulators for physical systems.  

Fortunately, it has been theoretically proven that GANs are capable of preserving all the constraints and statistics of the training data, up to the expressive capability of the generator and discriminator neural networks, if the global optimum is achieved in the training~\cite{goodfellow2014generative}. 
However, it is also well-known that traditional GANs has difficulty in convergence and lack robustness in the training~\cite{arjovsky2017wasserstein,radford2015unsupervised}.
Consequently, numerous efforts have been made to improve the stability and robustness in training GANs, leading to a number of GANs variants. For example, researchers proposed different loss functions and divergence formulations to remedy the vanishing gradients associated with the loss function used in the standard GANs~\cite{lin2017softmax, mao2017least,nowozin2016f,arjovsky2017wasserstein,gulrajani2017improved}. In contrast, the present work investigates the feasibility of using physical constraints to improve the training of GANs used to emulate physical systems. It is expected that our work will complement existing works in regularizing the training of physics-emulating GANs.

\subsection{Enforcing physical constraints in GANs}
\label{sec:intro-enforce}

The current work advocates using physical constraints to improve the robustness of training GANs. In our perspective, the physical constraints are considered regularization for the training of GANs to find the global minimum, and not merely as extra requirements to satisfy. We further demonstrate that even the imprecise constraint significantly improve the training of GANs. 
Such imprecise constraints are common in practical systems as they may arise from incomplete knowledge of the system properties or from reduced-order modeling of complex system dynamics.

Common constraints in physical systems stem from conservation laws, some of which can be incorporated into neural networks as penalty terms in the loss functions. However, practical systems in science and engineering often have properties that are only known partially, imprecisely, or with significant uncertainties~\cite{stinis2019enforcing}, which leads to imprecise physical constraints.  For example, a particle immersed in a fluid of unknown viscosity can be described by a momentum conservation equation with an uncertain dissipation term due to the viscous drag; a complex fluid of unknown constitutive relation can be described by the Navier--Stokes equations, but the stress term is inevitably imprecise. 
In addition to unknown system properties, imprecise constraints may also arise due to reduced-order modeling of complex systems. For example, the mean velocity and pressure fields of turbulent flows can be described by Reynolds averaged Navier--Stokes equations, which has a similar form to the Navier--Stokes equations but with a Reynolds stress term that needs to be modeled~\cite{pope00turbulent}. If one is tasked to generate mean flow fields, the RANS equations can serve as valuable constraints, although the governing equations now contain unknown terms and thus serve only as imprecise constraints. Enforcing imprecise physical constraints arising in GANs is the focus of the current paper.

\subsection{Scope and contributions of present work}

We propose a method to incorporate imprecise constraints into GANs and study the effects of such constraints on the training and generative performance of GANs. Compared to the recent work of Stinis et al.~\cite{stinis2019enforcing} studying physical constraints in GANs, the current work differs in two critical aspects: (i) that we focused on imprecise, conservation-law-like constraints and (ii) that the constraints are imposed on the generator. More specifically, Stinis et al.~\cite{stinis2019enforcing} evaluated the constraint residual and incorporated it as additional inputs of the discriminator, using the strength of discriminator to digest more information. In our work, we aim at directly strengthening the generator by enforcing the constraint upon the loss function associated with the generator. In terms of the imprecise constraints, small Gaussian random noises were taken as the constraint residual for true samples in~\cite{stinis2019enforcing}, which potentially allows small constraint residuals for the generated samples. However, it should be noted that the main reason of introducing small noises in~\cite{stinis2019enforcing} is not to enforce imprecise constraints upon generated samples but to avoid giving the discriminator even larger advantage over the relatively weaker generator in GANs setup~\cite{arjovsky2017towards}.
This work is complementary to the work presented in a companion paper that studied enforcing statistical constraints~\cite{wu2019enforcing}.
While precise constraints have previously been used as regularization in various fields (e.g., natural language processing~\cite{lee2019gradient}, lake temperature modeling~\cite{karpatne2017physics,read2019process}, general dynamical systems~\cite{stinis2019enforcing}, fluid flow simulations~\cite{magiera2020constraint,kim2019deep,eckert2018coupled,wiewel2020latent}, and more specifically in turbulent flow simulations and generation~\cite{mohan2020embedding,geneva2020multi,subramaniam2020turbulence}), the effects, performances, and best practices of imposing imprecise constraints in generative models still need further investigations.

Complex systems can be subject to two distinct types of constraints: deterministic constraints and statistical constraints. 
For example, the Navier--Stokes equations describing the motion and dynamics of incompressible fluid flows originally directly from the mass and momentum conservation,  which are referred to as \emph{deterministic constraints} in this paper. However, the solutions and properties of such PDEs exhibit sophisticated patterns and statistics due to the complex dynamics. For example, velocity increments in turbulent flows obey non-Gaussian distributions; the turbulent kinetic energy spectrum exhibit a decay rate of $-5/3$ in the universal range of the wavenumber space~\cite{pope00turbulent}. Such \emph{statistical constraints} describe the properties of an ensemble of system states rather than an individual state. 
Specifically, given an ensemble of instantaneous velocity and pressure fields of a turbulent flow, each sample must satisfy the deterministic constraints such as the divergence-free condition (mass conservation) and the momentum conservation. In addition, the ensemble as a whole must also satisfy the statistical constraints of turbulent flows (e.g., energy spectrum and velocity increment distribution).

Wu et al.~\cite{wu2019enforcing} investigated the effects of enforcing statistical constraints on GANs. It was found that enforcing statistical constraints improve the convergence rate and robustness (with respect to algorithmic parameters) in the training. Moreover, it was demonstrated that enforcing low-order statistics such as covariance caused GANs to generate samples that better conform to the high-order statistics as well.  In the current work, we focus on incorporating deterministic (as opposed to statistical) yet imprecise constraints into GANs and investigate the effects of such constraints.  

Generative models for fluid flows have a wide range of distinct applications scenarios, e.g., accelerating animation scene generation in computer graphics~\cite{ren2018visual} and providing inflow boundary conditions for turbulent flow simulations with direct simulations~\cite{subramaniam2020turbulence,mohan2020embedding}, large eddy simulations (LES) or hybrid LES/RANS simulations~\cite{xiao2012consistent,king2018deep}.
The present work is mainly motivated by current applications of GANs to generate inflow for LES and hybrid LES/RANS simulations. The quality of inflow turbulence plays a critical role on the performance of the LES. In fact, they are probably even more important than the sub-grid scale models in LES. Currently, such inflows are typically generated by precursor simulations, i.e., by performing an LES in a periodic domain, which can be computationally expensive and would inevitably introduce artificial periodicity. 
GANs also have potential applications to hybrid LES/RANS simulations, where turbulence fluctuations with specified mean field and statistics must be generated in the grey areas. A grey area refers to an LES region that is located immediately downstream of a RANS region and the simulated flows do not have sufficient instability to become turbulence.

This rest of the paper is organized as follows. In Section~\ref{sec:method}, a brief overview of GANs is given,  and the proposed methodology of imposing physical constraints is presented. In Section~\ref{sec:results}, a test case with simple geometric constraints and another test case of fluid flows with differential constraints are presented to demonstrate the merit of the constrained GANs. Finally, Section~\ref{sec:conclusion} concludes the paper.

\section{Methodology}
\label{sec:method}
Generative adversarial networks use a two-player competitive game strategy to train a generator that maps a generic (e.g., Gaussian) distribution in the latent space to a distribution in the data space~\cite{goodfellow2014generative}. Once trained, the generator can produce realistic examples conforming to the data distribution by drawing samples from the generic distribution. The current work focuses on enforcing deterministic, conservation-law-like physical constraints on the generated samples. In this section, we first describe the general idea of GANs architecture and their training procedure. Then, we introduce a generic representation of imprecise physical constraints and explain why they are enforced on the loss function of the generator. Finally, we discuss a few practical implementation issues in stabilizing the training.

\subsection{Architecture and training of GANs}
\label{sec:method-gans}

\begin{figure}
    \centering
  \includegraphics[width=0.95\textwidth]{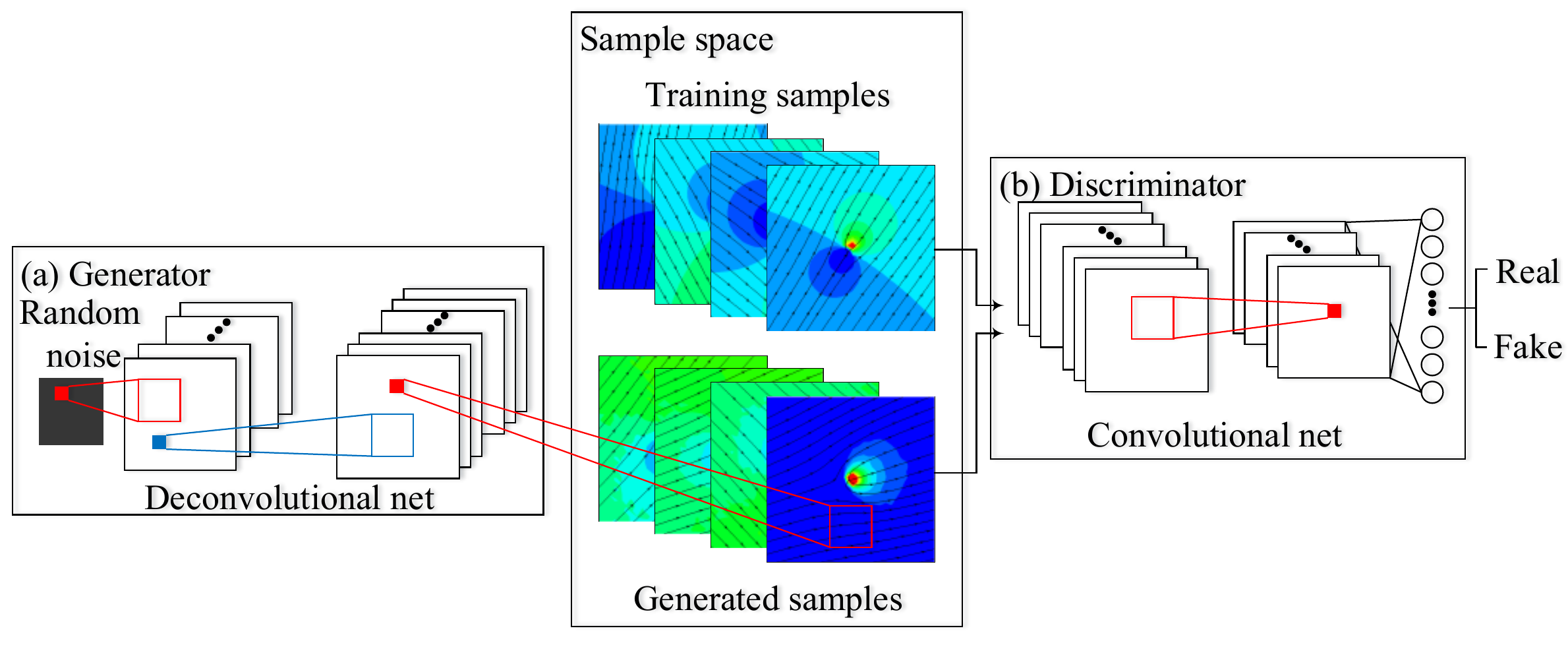}
    \caption{Architecture of the generative adversarial networks (GANs). The constrained GANs architecture modifies the standard GANs  by adding a penalty term to the loss function of the generator to impose physical constraints, which is detailed in Fig.~\ref{fig:graphic-abstract}.  
    }
    \label{fig:cons-cGAN}
\end{figure}

Generative adversarial networks (GANs) were first proposed by Goodfellow et al.~\cite{goodfellow2014generative}, which features a novel neural network architecture consisting of two competing networks: a generator $\bm{G}$ and a discriminator $\bm{D}$. The loss function of standard GANs is:
\begin{equation}
    \min_{\boldsymbol{G}} \max_{\boldsymbol{D}} V(\boldsymbol{D},\boldsymbol{G}) = \mathbb{E}_{X\sim p_\text{data}(X)}[\log \boldsymbol{D}(X)] + \mathbb{E}_{Z\sim p_{z}(Z)}[\log (1-\boldsymbol{D}(\boldsymbol{G}(Z))]
    \label{eq:lossfunction-GANs}
\end{equation}
where $p_\textrm{data}(X)$ and $p_z(Z)$ denote the probability distributions of training data and latent space vector $Z$, respectively, and $\mathbb{E}$ indicates expectation.
The generator $\bm{G}$ takes a vector $Z$ in the latent space and maps it to an element in the data space. For example, for the task of flow field synthesis, $Z$ can be a vector drawn from Gaussian or uniform distributions, which is mapped to the space of flow fields as illustrated in Fig.~\ref{fig:cons-cGAN}. The discriminator~$\bm{D}$ takes a sample $X$ (either a training sample or a generated sample) in the data space as input and outputs a label to indicate whether the input $X$ is drawn from the training data (1 for true and 0 for false).

The training of GANs involves solving the ``min--max'' optimization problem, where the generator and the discriminator compete with each other to achieve a Nash equilibrium~\cite{goodfellow2014generative}. Specifically, the discriminator $\bm{D}$ is first optimized (with $\bm{G}$ kept fixed) to correctly distinguishes a synthesized sample by generator from a sample from the training data as accurate as possible (i.e., $\bm{D}(\bm{G}(Z)) = 0$ and $\bm{D}(X) = 1$), which maximizes both expectations in the loss function. Then, the generator is optimized (with $\bm{D}$ kept fixed) to generate a sample as realistic as possible in the viewpoint of the discriminator. The objective is to trick the discriminator to consider the sample drawn from the training data (i.e., $\bm{D}(\bm{G}(Z))=1$) as much as possible, which minimizes the expectation of $\log (1-\boldsymbol{D}(\boldsymbol{G}(Z))$ in the loss function. This two-steps process is repeated until the discriminator correctly distinguishes generated samples from training samples with a probability of 50\%, where the Nash equilibrium is achieved. Intuitively, in this scenario the generated samples are realistic enough that the discriminator is completely incapable of distinguishing them from training samples.

Goodfellow et al.~\cite{goodfellow2014generative} proved that the global optimum of the ``min-max'' problem in Eq.~\eqref{eq:lossfunction-GANs} is achieved if and only if the generator can exactly mimic the probability distribution of the training data, i.e., $p_g = p_\textrm{data}$. Such a statement builds upon the proof that, with a fixed generator $\bm{G}$, the solution
\begin{equation}
    \boldsymbol{D^*}_{\boldsymbol{G}}(X)=\frac{p_\text{data}(X)}{p_\text{data}(X)+p_\text{g}(X)}
    \label{eq:optimized_D}
\end{equation}
minimize the loss function in Eq.~\eqref{eq:lossfunction-GANs}. That is, the Nash equilibrium $\bm{D}(X) = 0.5$ is achieved if and only if $p_g = p_\textrm{data}$. That is, under Nash equilibrium the generator is guaranteed to reproduce all the statistics of the training data, including statistical moments and correlations of any order.

While the theoretical capability of GANs in reproducing data statistics is reassuring, the ideal scenario requires the global optimum is achieved in the training. Unfortunately, 
standard GANs are difficult to train. One of the reasons is that the gradient vanishes as the discriminator $\bm{D}$ approaches the optimum, which prevents the generator $\bm{G}$ being further optimized. As it is unlikely that both the discriminator and the generator would achieve optimum simultaneously, the vanishing gradient can lead to an unsatisfactory generator.  

In light of the vanishing gradient problem of the standard loss function, numerous researchers have proposed alternative loss functions to improve the training of GANs. Arjovsky et al.~\cite{arjovsky2017wasserstein} used the Earth-Mover (Wasserstein-1) distance $W(q, p)$, with which non-zero gradient of the loss function still exists even with an optimal discriminator $\bm{D}$. The corresponding loss function is:
\begin{equation}
    \min_{\boldsymbol{G}} \max_{\boldsymbol{D} \in \Omega} V_\textrm{WGAN}(\boldsymbol{D},\boldsymbol{G}) = \mathbb{E}_{X\sim p_\text{data}(X)}[\boldsymbol{D}(X)] - \mathbb{E}_{z\sim p_\text{z}(z)}[ \boldsymbol{D}(\boldsymbol{G}(z))]
    \label{eq:lossfunction-WGANs}
\end{equation}
where $\Omega$ denotes the set of all 1-Lipschitz functions.
The network weights of generator $G$ are clipped within a compact space to satisfies Lipschitz constraint empirically. Gulrajani et al.~\cite{gulrajani2017improved} pointed out that a poorly designed compact space would result in vanishing or exploding gradient, and they proposed using a gradient penalty (GP) term to enforce the Lipschitz constraint. The new loss function is 
\begin{equation}
    V_\textrm{WGAN-GP}(\boldsymbol{D},\boldsymbol{G}) = V_\textrm{WGAN}(\boldsymbol{D},\boldsymbol{G}) + \beta \mathbb{E}_{\hat{X}\sim P(\hat{X})}\left[\left(\Vert \nabla_{\hat{X}}\boldsymbol{D}(\hat{X})\Vert_2-1\right)^2\right]
    \label{eq:lossfunction-WGAN-GP}
\end{equation}
where $\hat{X}$ is the sample uniformly generated between the generated data $\boldsymbol{G}(z)$ and the training sample $X$, and $\beta$ is the weight of gradient penalty. In this work we used WGAN-GP to achieve improved stability in the training.

In many applications, it is desirable to impose some conditions on the generated samples, e.g., generating velocity fields that correspond to specified boundary conditions or conform to specified statistics. Such conditioning can be achieved by augmenting the latent space vector $Z$ in the standard GANs with an auxiliary variable $\zeta$ that encodes the specified conditions. This modification leads to an architecture referred to as \emph{conditional GANs} (cGANs)~\cite{mirza2014conditional}. The loss function in Eq.~\eqref{eq:lossfunction-GANs} for standard GANs can be modified for conditional GANs as follows:
\begin{equation}
    \min_{\boldsymbol{G}} \max_{\boldsymbol{D}} V(\boldsymbol{D},\boldsymbol{G}) = \mathbb{E}_{X\sim p_\text{data}(X)}[\log \boldsymbol{D}(X,\zeta)] + \mathbb{E}_{Z\sim p_\text{z}(Z)}[\log (1-\boldsymbol{D}(\boldsymbol{G}(Z,\zeta))] ,
    \label{eq:lossfunction-cGANs}
\end{equation}
where random vectors $X$ and $Z$ conform to their respective distributions while $\zeta$ is a deterministic condition given to the generator. For example, if we would like the GANs to generate a specific hand-written digit (e.g., 8) rather than an arbitrary digit, then $\zeta$ shall be specified to the label ``8''.  Here, both the generator and the discriminator explicitly depends on the auxiliary variables $\zeta$ encoding the specified conditions. Subsequently, when using cGANs to generate samples, the conditioning-encoding vector $\zeta$ and the latent space vector $Z$ are both provided as inputs to the generator $G$. This causes the generator to produce samples that both conform to the data distribution and satisfy the specified conditions. Recently, cGANs have been successfully used to generate solutions for PDEs with specified boundary conditions describing heat diffusion and fluid flows problems~\cite{farimani2017deep}. It can be seen from Eqs.~\eqref{eq:lossfunction-GANs} and~\eqref{eq:lossfunction-cGANs} that the GANs and cGANs formulations are similar. Therefore, it is straightforward to extend the definitions of the GANs variants in Eqs.~\eqref{eq:lossfunction-WGANs} and~\eqref{eq:lossfunction-WGAN-GP} to the corresponding cGANs. The extensions are omitted here for brevity.

\subsection{Representation of generic physical constraints}
\label{sec:method-physics}
Deterministic physical constraints exhibit themselves in a wide range of forms from simple algebraic equation to nonlinear integro-differential equations and inequalities. 
First, for simpler systems the conservation laws can be directly expressed as algebraic expression.  For example, the conservation of potential energy (PE) and kinetic energy (KE) of an idealized pendulum is simply written as $\text{KE} + \text{PE} = \text{constant}$; the energy conservation of inviscid, irrotational flows is $p + \frac{1}{2} \rho |\bm{v}|^2  + \rho g h = \text{constant}$, where $p$ is the pressure, $|\bm{v}|$ is the velocity magnitude,  $\rho$ is density, $g$ is the gravitational acceleration, and $h$ is elevation, respectively.
Second, the most common physical principles such as conservation laws take the form of differential (sometimes integro-differential) equations $\mathcal{N}[u] = f$, where $\mathcal{N}[\cdot]$ denotes differential operators. Common examples in computational mechanics include Navier--Stokes equations describing mass and momentum equations of fluid flows and equations of linear elasticity describing the force balance of solids. Many constraints can be transformed to linear systems of the following form via linearization and numerical discretization: 
\begin{equation}
    \mathsf{N}  
\mathbf{u}  =  
\mathbf{f} \, , 
\label{eq:linear}
\end{equation}
where $\mathsf{N}$ is the matrix resulted from the discretization of differential operator $\mathcal{N}$, $\mathbf{u}$ is the vector of the field to be solved for (e.g., velocity or displacement), and $\mathbf{f}$ is the forcing vector.  Finally, some physics laws take the form of inequality equations, possibly involving differential operators. For example, the second law of thermodynamics states that the total entropy
of a closed system can only increase or stays the same, i.e.,
$dS \ge \delta Q/T$ where $dS$, $\delta Q$, and $T$ denotes entropy increment, heat addition, and surrounding temperature, respectively. 

In view of the diverse examples physical laws enumerated above, we write the constraints in a generic form as follows~\cite{karpatne2017physics}:
\begin{equation}
\mathcal{H}[\mathbf{u}] \le 0 \, ,
\label{eq:generic-constraint}
\end{equation}
where $\mathcal{H}$ denotes an algebraic operator, which can involve nonlinearity and can be obtained from numerical discretization of the integro-differential operators in the original physics constraints.
Typically, $\mathcal{H}$ is chosen to be a non-negative function such as norm, so requiring $\mathcal{H}[\mathbf{u}] \le 0$ would effectively lead to $\mathcal{H}[\mathbf{u}] = 0$.
As an example, the linear equation~\eqref{eq:linear} above can be cast into an inequality as
$\| \mathsf{N}  \mathbf{u} - \mathbf{f}  \|^2 \le 0$,
where $\|\cdot\|$ indicates the Euclidean vector norm. 

\subsection{Embedding constraints into the generator of GANs}
\label{sec:method-embed}
\begin{figure}
    \centering
    \includegraphics[width=0.95\textwidth]{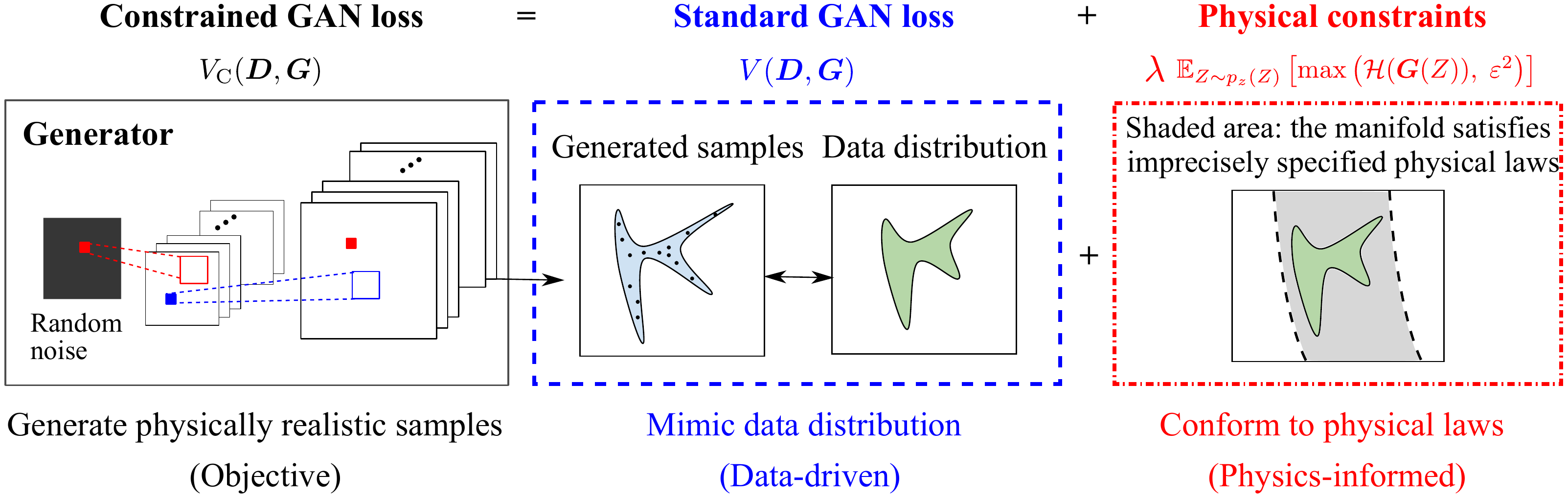}
    \caption{
    Schematic illustration of the physics-constrained GAN proposed in the current work, which generates physically realistic samples by combining two sources of knowledge: (1) learning from training data distribution and (2) enforcing imprecise physical constraints. The blue part (middle panel) stems from standard GANs, and the red part (right panel) indicates contributions of current work.}
    \label{fig:graphic-abstract}
\end{figure}

In order to impose deterministic physical constraints in GANs, in this work we add a penalty term to the loss function of the generator based on the generic representation of constraints in Eq.~\eqref{eq:generic-constraint}. That leads to the following loss function for physics-constrained GANs:
\begin{align}
    V_\textrm{C}(\boldsymbol{D},\boldsymbol{G}) 
    & = V(\boldsymbol{D},\boldsymbol{G}) + \lambda C_\textrm{phys} \label{eq:loss-PIGANs} \\
    \quad \textrm{with} \quad
    C_\textrm{phys} & = \mathbb{E}_{Z\sim p_{z}(Z)}\left[\max\left(\mathcal{H}(\boldsymbol{G}(Z)), \; 0\right)\right] ,
\end{align}
where $\lambda$ denotes the penalty coefficient, and $V$ is the loss function of the baseline GANs to be constrained.  As the physical constraint term is independent of the discriminator $D$, its gradient is non-zero even if the discriminator $D$ is close to optimum. Therefore, adding physical constraints to the generator loss function provides an alternative approach for combating the vanishing gradient problem in the training of GANs.
Note that Eq.~\eqref{eq:loss-PIGANs} is generally applicable to different variants of GANs. The general concept of physics-constrained GAN is illustrated in Fig.~\ref{fig:graphic-abstract}, with the imprecise physical constraints explained in Sec.~\ref{sec:method-implement} in more details.
The coefficient~$\lambda$ must be chosen to conform to the data distribution and to satisfy the physical constraint. We will discuss the choice of $\lambda$ in Section~\ref{sec:res-discuss} with a comprehensive parametric study.

The physical constraints are better imposed on the generator rather than on the discriminator probably because such constraints help the weaker party (generator) in the two-player game, which in turn accelerates the convergence to Nash equilibrium between the generator and the discriminator during the training of GANs. In the two-player game, the gradient vanishes for the standard GANs if the discriminator achieves its optimum first, and thus the generator would not be able to further update its weights. In this situation, the physical constraint introduces an additional gradient and makes it possible to train the generator further.  Therefore, embedding the physical constraints into the generator helps the generator and the discriminator achieve equilibrium more synchronously. However, the additional constraint may lead to an optimum that is different from the original unconstrained loss function as can be seen in Eq.~\eqref{eq:loss-PIGANs}.

\subsection{Implementation considerations and imprecise constraints}

\label{sec:method-implement}

In many practical applications, the physical knowledge is commonly limited, i.e., the physical constraints can only be described imprecisely instead of being specified exactly. This type of physical constraints demands a modified definition of physics-constrained term $C_\textrm{phys}$ accordingly. In this work, such imprecise constraints are defined as
\begin{equation}
    C^{\varepsilon}_\textrm{phys} = \mathbb{E}_{z\sim p_\text{z}(Z)}\left[\max\left(\mathcal{H}(G(Z)), \; \varepsilon^2 \right)\right]
    \label{eq:soft-cons}
\end{equation}
where $\varepsilon^2 \geq 0$ is threshold that denotes the square of the uncertainty interval on the corresponding physical constraints. With such a definition, we enforce an imprecise constraint $\mathcal{H}(G(Z))\leq \varepsilon^2$ when training the constrained GANs.

In order to facilitate training of the constrained GANs, it is desirable to keep the penalty terms related to the physical constraint of the same order of magnitude as the  term $V(\bm{G}, \bm{D})$ associated with the baseline GANs. To this end, the term $\mathcal{H}(\boldsymbol{G}(Z))$ in the loss function (Eq.~\eqref{eq:soft-cons}) is further replaced by 
\begin{equation}
   \log(\mathcal{H}(\boldsymbol{G}(Z)) + 1)
    \label{eq:loss-log}
\end{equation}
in our implementation.
Since $\mathcal{H}$ is a non-negative function by construction,  the penalty term $\log(\mathcal{H}(\boldsymbol{G}(Z)) + 1)$ above is guaranteed to be positive (and thus ensured to be a penalty rather than a reward).
This modification is proposed to address two practical difficulties. First, in the initial phase of the training, the generated data often have dramatic departure from the physical constraints, leading to very large values of the loss term $C_\text{phys}$. Such a disparity in the loss function where one completely dominates the other will lead to difficulty in training.
Second, in the final phase of the training, the physical constraint related loss function is much smaller than the standard GANs value term. This makes it difficult to enforce the physical constraint. By using the logarithm of the physical constraint penalty, the two difficulties above can be addressed simultaneously. Note that the constant $1$ is added to ensure the strict positivity of the logarithm and to avoid singularity (i.e., $\log 0$).

The physics-constrained GANs are implemented based on the open-source software library  \verb+TensorFlow+~\cite{abadi2016tensorflow}. Our implementation is built upon the standard, baseline GANs implementation by Kristiadi et al.~\cite{kristiadi2019collection}. The source code for the physics-constrained GANs and the example cases presented in Section~\ref{sec:results} below are publicly available in a GitHub repository~\cite{zeng2019enforcing}.

\section{Results}
\label{sec:results}

In this section we use two cases to demonstrate the performance of the proposed method to constrain GANs and to investigate the effects of such enforcing constraints. The constraints are chosen to be representative of, yet much simpler than, the physical constraints in sciences and engineering. 
Consider the example scenario of using GANs to generate velocity fields of turbulent flows on a mesh of $100 \times 100 \times 100$ grid points. Apparently, each instantaneous flow field (snapshot) consists of $3\times10^6$ degrees of freedom, which is analogous to a picture of $10^6$ pixels with three channels (corresponding to the three velocity components here) in each pixel. However, the number of intrinsic degrees of freedom is much smaller, because the velocities at the grid points are not completely independent. For example, the divergence of the velocity must be zero everywhere. Moreover, the turbulent kinetic energy distribution at different wavenumbers also determines the smoothness of the field, which further constrains the intrinsic degrees of freedom.

Inspired by the example of turbulent flows above, we propose the following two test cases. The first case has a geometrical constraint. We use GANs to generate circles, each of which is represented by 100 points evenly distributed in the azimuthal direction. As each point is determined by its $x$- and $y$-coordinates, the apparent degree of freedom of each sample (a circle) is 200 for GANs. However, intrinsically each sample has only one degree of freedom, i.e., its radius. The simple example is convenient to visualize and analyze, but it has a strong constraint that is representative of those in many physical applications. The second case has a differential constraint stemming from the divergence-free condition of the velocity field. The training samples consist of velocity fields on a $32\times32$ mesh generated from a complex potential parameterized by three parameters. Again, the underlying physics forms a strong constraint, leading to a dramatic difference between the apparent and intrinsic degrees of freedom.

\subsection{Geometrical constraint: generating circles}

Samples for the first case are two-dimensional circles on the $x$--$y$ plane with different radii.
The main objectives include (i) illustrating the merit of incorporating known constraints into the generator and (ii) demonstrating the effects of incorporating imprecise constraints as proposed in Eq.~\eqref{eq:soft-cons}.

\subsubsection{Problem description}
\label{sec:problem-description}

\begin{figure}
    \centering
    \includegraphics[width=0.7\textwidth]{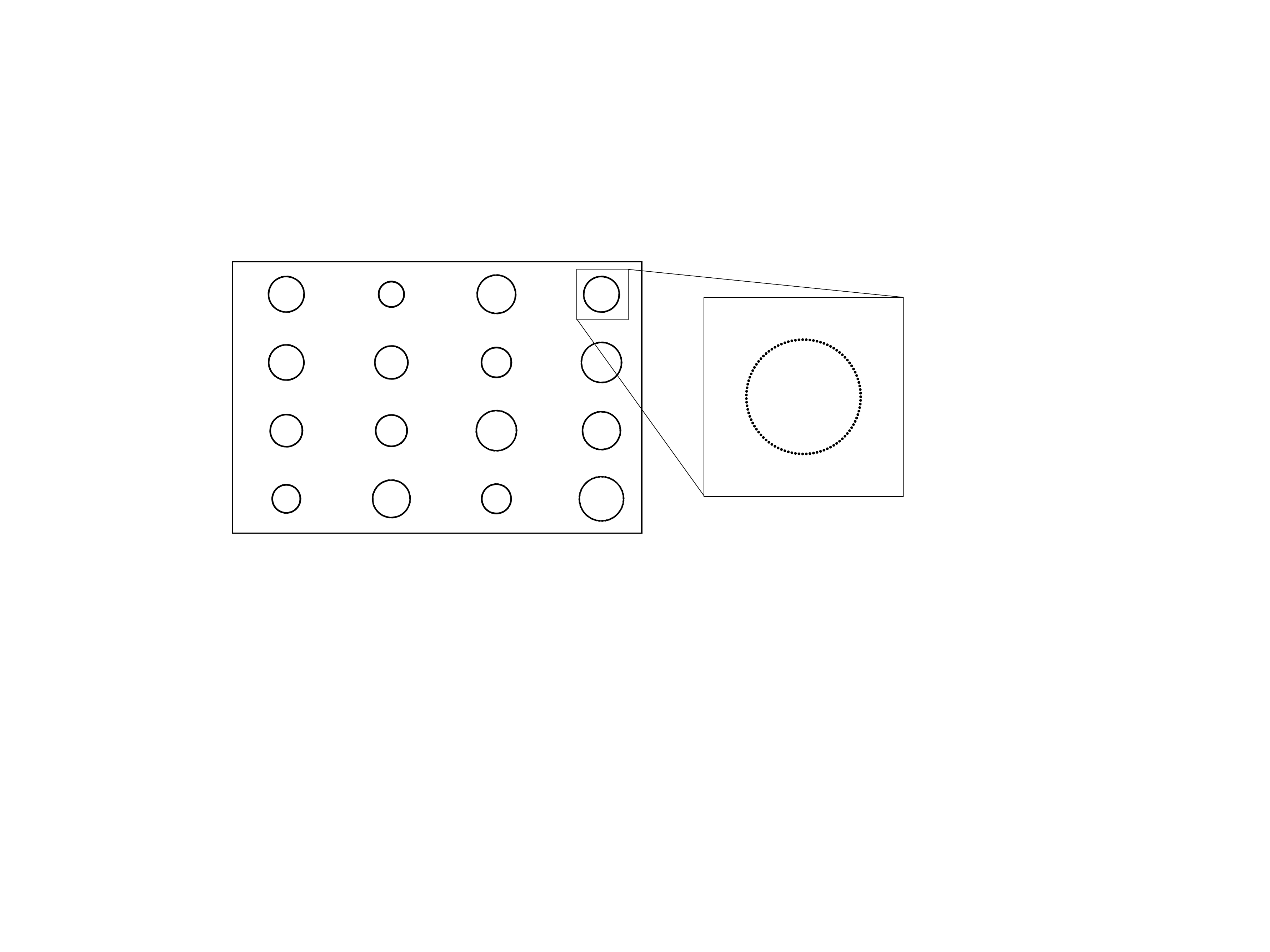}
    \caption{
    Representative examples from the training samples in the geometric constraint case. Each sample consists of 100 points as is evident from the zoom-in view in the right panel (albeit appearing as solid lines above due to compact spacing in the left panel). These points are evenly distributed along the circumferential direction of a circle. The radius $r$ of each circle is randomly drawn from a uniform distribution in the range $[0.4, 0.8]$.}
    \label{fig:M-traing-samples}
\end{figure}

\begin{figure}
    \centering
    \includegraphics[width=0.42\textwidth]{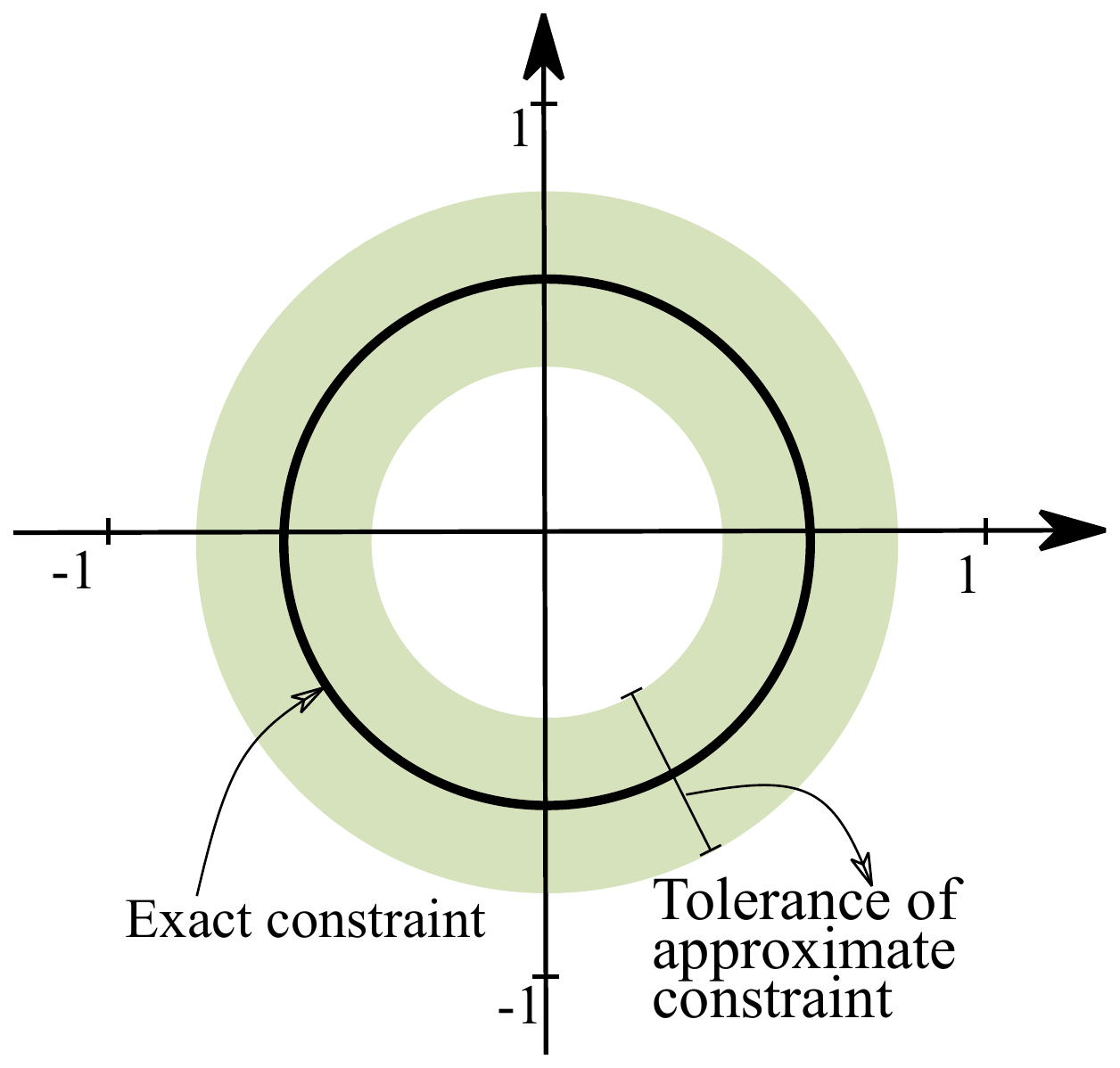}
    \caption{
    Schematic illustration of the imprecise constraint. The constraint described in Eq.~\eqref{eq:soft-circle} for generating a circle is used as an example. The radius $r$ associated with the exact constraint is indicated as the solid circle. The tolerance $\varepsilon$ of the constraint is indicated as annulus-shaped shade.
    The loss function associated with the imprecise constraint activated only when a point falls outside the shade. }
    \label{fig:circle-constraints}
\end{figure}

The training samples are the circles with different radii. The parametric constrained function of circles is given as
\begin{equation}
\begin{aligned}
 \left\{ \begin{array}{rcr}
   x = r\cos(\psi)\\
   y = r\sin(\psi)
   \end{array} \right. \\
\end{aligned}
\label{eq:circle}
\end{equation}
where $\psi$ is the angle in polar coordinate, and $r$ is the radius. $\bm{x}^{(i)}=[x^{(i)}, y^{(i)}]^\top$ denotes the points on the circle in Cartesian coordinates, where $i \in \{1,2,...,100\}$ is the index of point of a generated sample. The training samples consist of $N=5000$ circles whose radius is drawn from a uniform distribution between 0.4 and 0.8. Some examples are shown in Fig.~\ref{fig:M-traing-samples}.
The goal is to first train GANs with the training dataset of circles and then use the trained GANs to generate samples. We train standard GANs, standard cGANs, and constrained cGANs proposed here with the same settings as presented in Table~\ref{tab:GAN-settings}. The imprecise constraint is defined as follow:
\begin{equation}
    C^\textrm{approx} = \mathbb{E}_{Z\sim p_{z}(Z)}\left[ \sum_{i} \max\left(\left( \|\bm{x}^{(i)}\| - \zeta \right)^2, \; \varepsilon^2 \right)\right] ,
    \label{eq:soft-circle}
\end{equation}
where $\|\cdot\|$ indicate Euclidean norm, $\zeta$ denotes the specified radius, and $\varepsilon$ denotes the tolerance of the specified constraint. The imprecise constraint is schematically illustrated in Fig.~\ref{fig:circle-constraints}.

\begin{table}[!htb]
\caption{Detailed settings of GANs architecture and their training in the geometric constraint case.}
\centering
\begin{tabular}{P{6cm} P{6cm}}
\hline
Loss function for baseline GAN & WGAN-GP \\
Dimension of latent space & $Z\in\mathbb{R}^{30}$ \\
Activation functions & LeakyReLU$^{\text{(a)}}$ and tanh$^{\text{(b)}}$ \\
Epochs & up to 200\\
\hline
\end{tabular}
\begin{flushleft}
\small
Notes: (a) LeakyReLU \cite{maas2013rectifier} activation function is used for both the generator and discriminator except the last layer.
The slope of the LeakyReLU activation function is 0.2 for negative inputs.\\
(b) The tanh activate function is used in the last layer of the generator to normalize the output of the generator within the range~$[-1,1]$.
\end{flushleft}
\label{tab:GAN-settings}
\end{table}

In order to evaluate and compare the performances of different GANs, three metrics are used to evaluate the generated samples. The first metric is the magnitude of the original loss function, which shows the convergence of different GANs in training. 
The second metric is the deviation of generated samples from the circle with an optimal radius obtained from least-square fitting, which is defined as follow:
\begin{equation}
\mathcal{E}_\text{dev} = \frac{1}{n} \sum_{i=1}^n\left|(x_i^2 + y_i^2) - \bar{r}^2 \right|
\label{eq:M-deviation}
\end{equation}
where $\left| \cdot \right|$ indicates absolute value, $n=100$ is the number of points used to represent the circle, and $\bar{r}$ is the radius obtained through least-square fitting of all 100 points in the sample.
To assess the overall deviation of the generated samples, we compute the sample mean $\langle \mathcal{E}_\text{div} \rangle$ by averaging that of all generates samples, where $\langle \cdot \rangle$ indicates ensemble averaging. For standard cGANs and constrained cGANs, $\bar{r}$ can be regarded as the estimate of auxiliary variable~$\zeta$, i.e., the radius in this case. There will be a bias between the estimated radius $\bar{r}$ and the given radius $\zeta$, and we further define the bias as another metric based on the deviation from the specified radius:
\begin{equation}
\mathcal{E}_\text{bias}
= \vert \bar{r} - \zeta \vert \; ,
\label{eq:M-bias}
\end{equation}
Sampling averaging is performed similarly as above to obtain $\langle \mathcal{E}_\text{bias} \rangle$.

We remark that finding a radius $\bar{r}$ from a collection of 100 points does not create a self-fulling prophecy. It is true that any 100 points would give a $\bar{r}$ value via the least square fitting procedure above. However, if the generated image is deviate from a circular shape, the deviation metric $\mathcal{E}_\textrm{dev}$ as defined above would be very large. Only when all the 100 points more or less fall onto a circle would the deviation $\mathcal{E}_\textrm{dev}$ be small, which justifies the deviation metric above.

\subsubsection{Results and discussions}
\begin{figure}[htb]
    \centering
    \begin{subfigure}[b]{0.49\textwidth}
        \includegraphics[width=\textwidth]{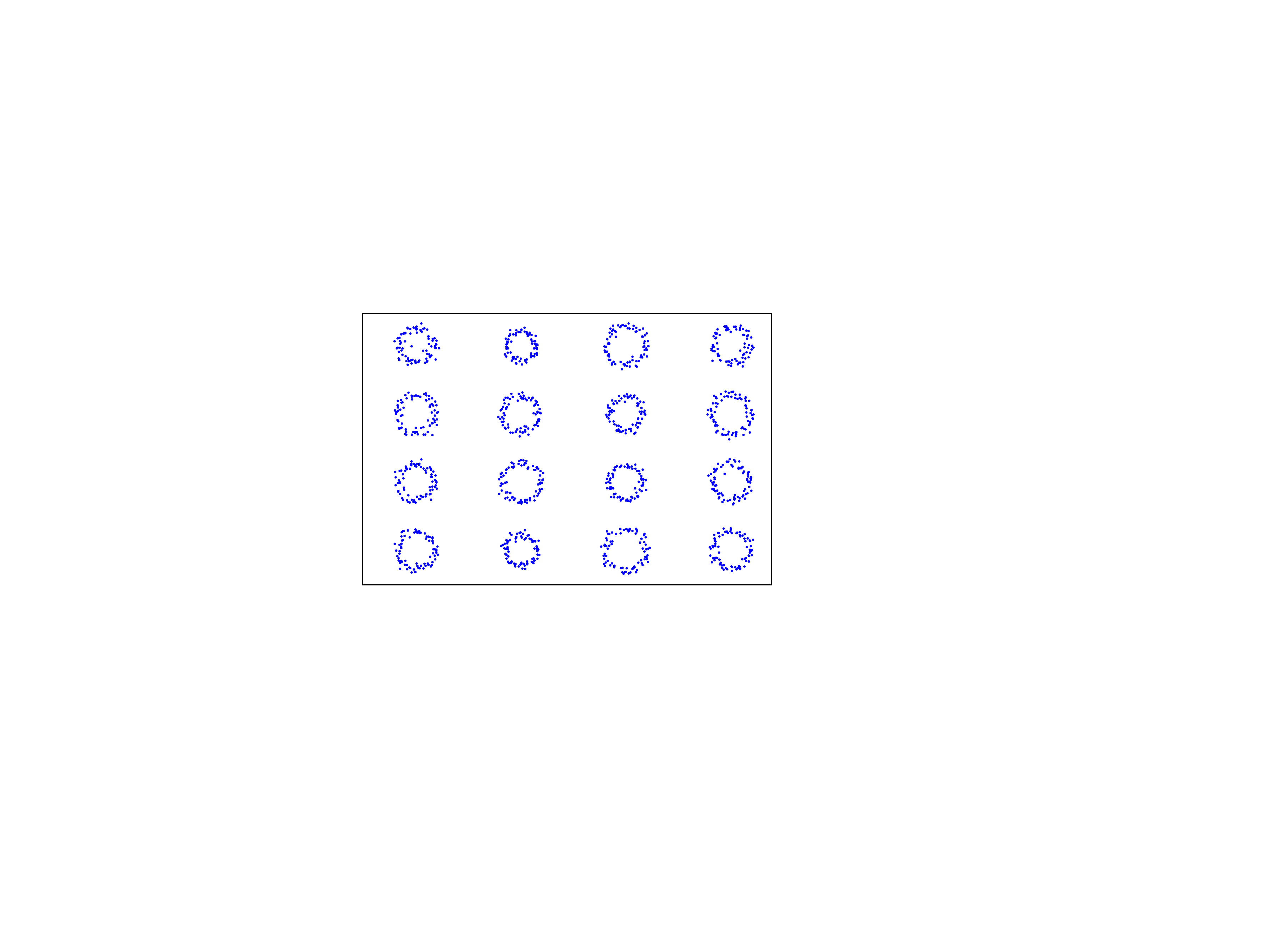}
        \caption{Standard GANs}
    \end{subfigure}\\
    \begin{subfigure}[b]{0.49\textwidth}   
        \includegraphics[width=\textwidth]{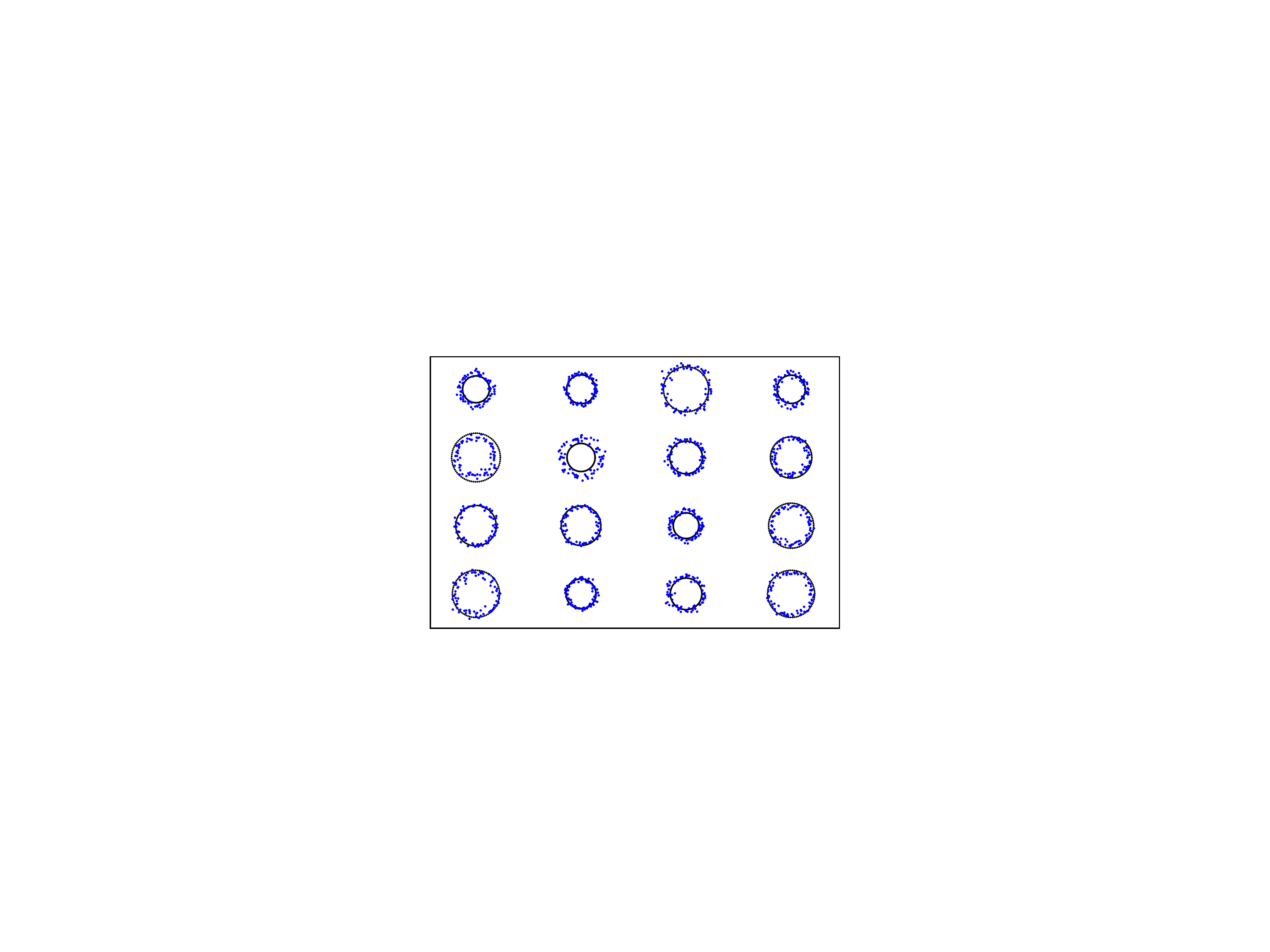}
        \caption{Standard cGANs}
    \end{subfigure}
    \begin{subfigure}[b]{0.49\textwidth}
        \includegraphics[width=\textwidth]{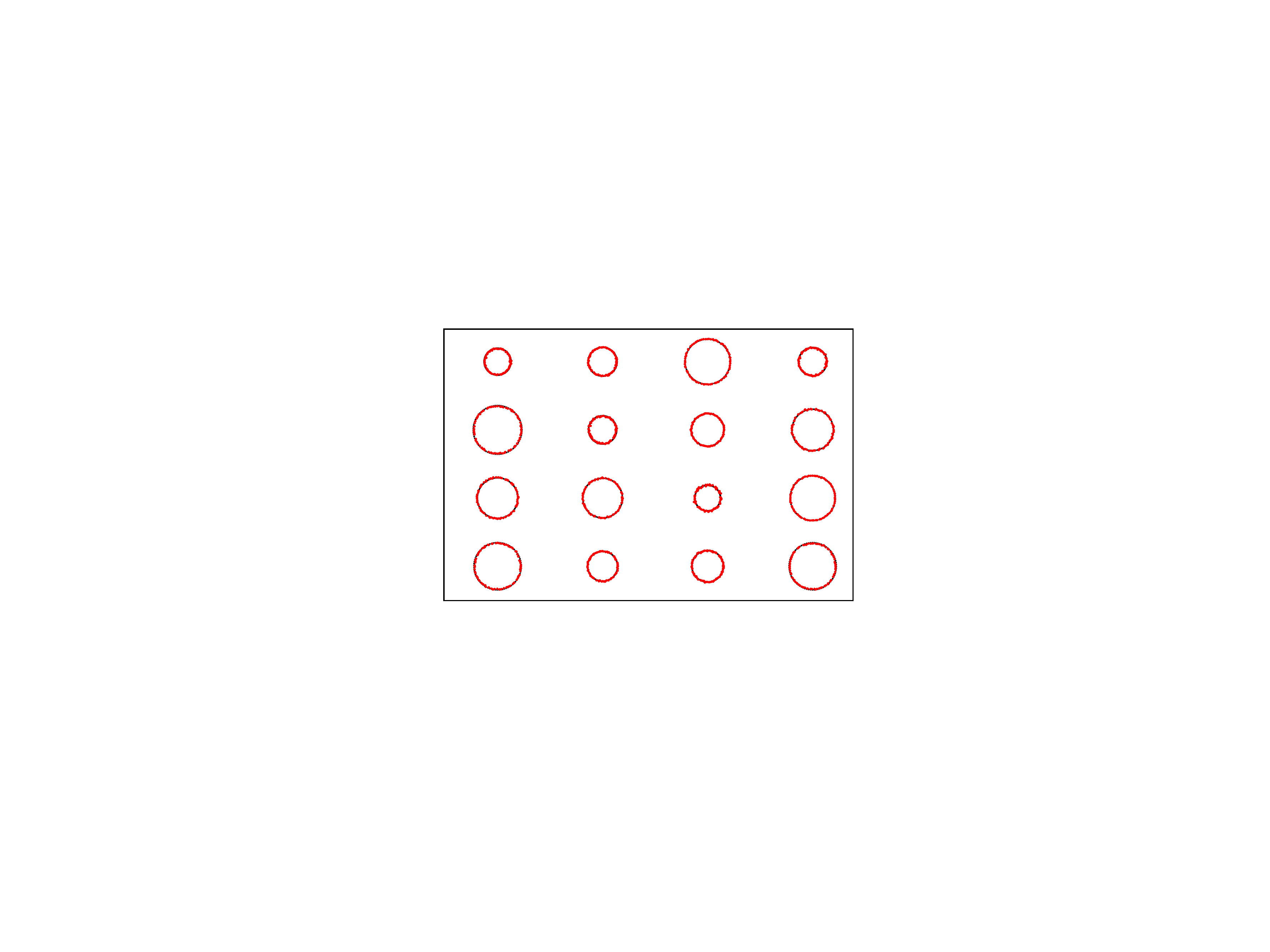}
        \caption{Constrained cGANs}
    \end{subfigure}
    \caption{
    Sample circles (images consisting of 100 dots) generated by (a) standard GANs, (b) standard cGANs, and (c) constrained cGANs. 
    The standard GANs are tasked to generate circles of any radius, while the standard and constrained cGANs are tasked to generate circles of specified radius, denoted as solid lines in panels (b) and (c). In panel (c), the generated (red/gray) dots fall on top of the solid circles of specified radius and are thus difficult to discern.}
    \label{fig:Generated-data}
\end{figure}

We first set $\varepsilon=0$ to test the performance of the baseline GANs with exact constraint, including both the standard GANs and standard cGANs. Specifically, 5000 circles are generated by using standard GANs, standard cGANs and constrained cGANs, with some generated examples shown in Fig.~\ref{fig:Generated-data}. By training on perfect circles with different radius, it can be seen in Fig.~\ref{fig:Generated-data}a that even standard GANs can generate circle-like samples, although noticeable noises exist in the generated samples. 

Noted that standard GANs can only generate samples randomly, i.e., the radius of the generated sample cannot be controlled in this simple test case. In order to generated samples with specified conditions, cGANs are also studied in this work. The solid circles represent the truth with the corresponding radius as the condition label in Figs.~\ref{fig:Generated-data}b and~\ref{fig:Generated-data}c. It can be seen that the constrained cGAN generates samples that have much better agreement with the truth as compared to the standard cGANs. Moreover, although the noise level of standard cGANs results is slightly lower than that in standard GANs, a clear deviation from perfect circle can still be seen in the generated samples from standard cGAN in Fig.\ref{fig:Generated-data}b. In contrast, the generated samples from the constrained cGANs demonstrate a nearly perfect agreement with the truth, confirming the merit of introducing constraints to GANs.

\begin{figure}[!htb]
    \centering
    \includegraphics[width=0.4\textwidth]{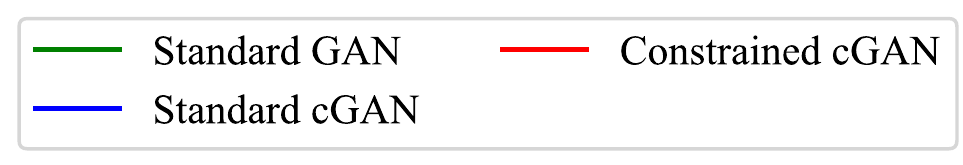} \\
    \begin{subfigure}[b]{0.44\textwidth}
        \includegraphics[width=\textwidth]{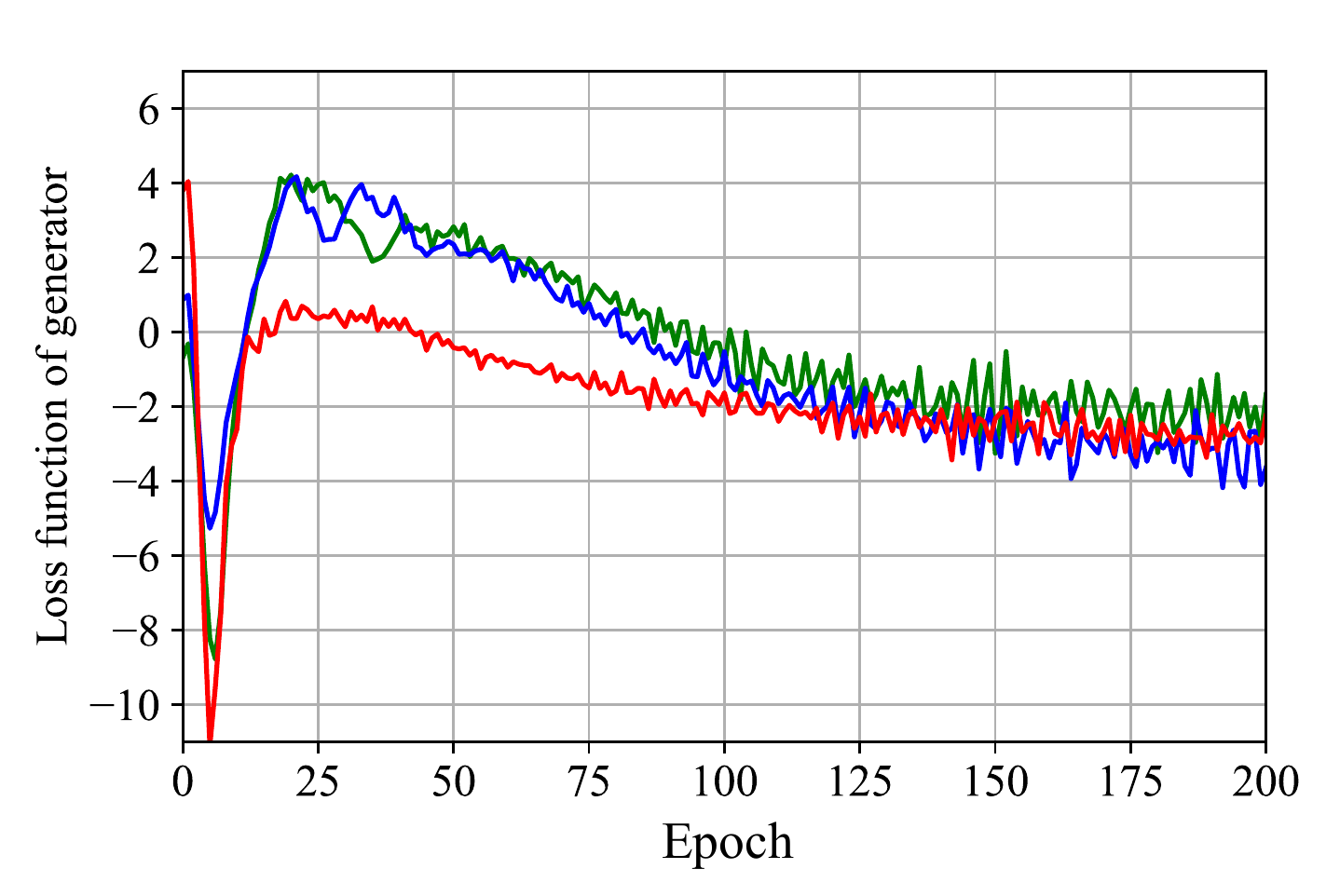}
        \caption{Generator}
    \end{subfigure}
    \begin{subfigure}[b]{0.44\textwidth}   
        \includegraphics[width=\textwidth]{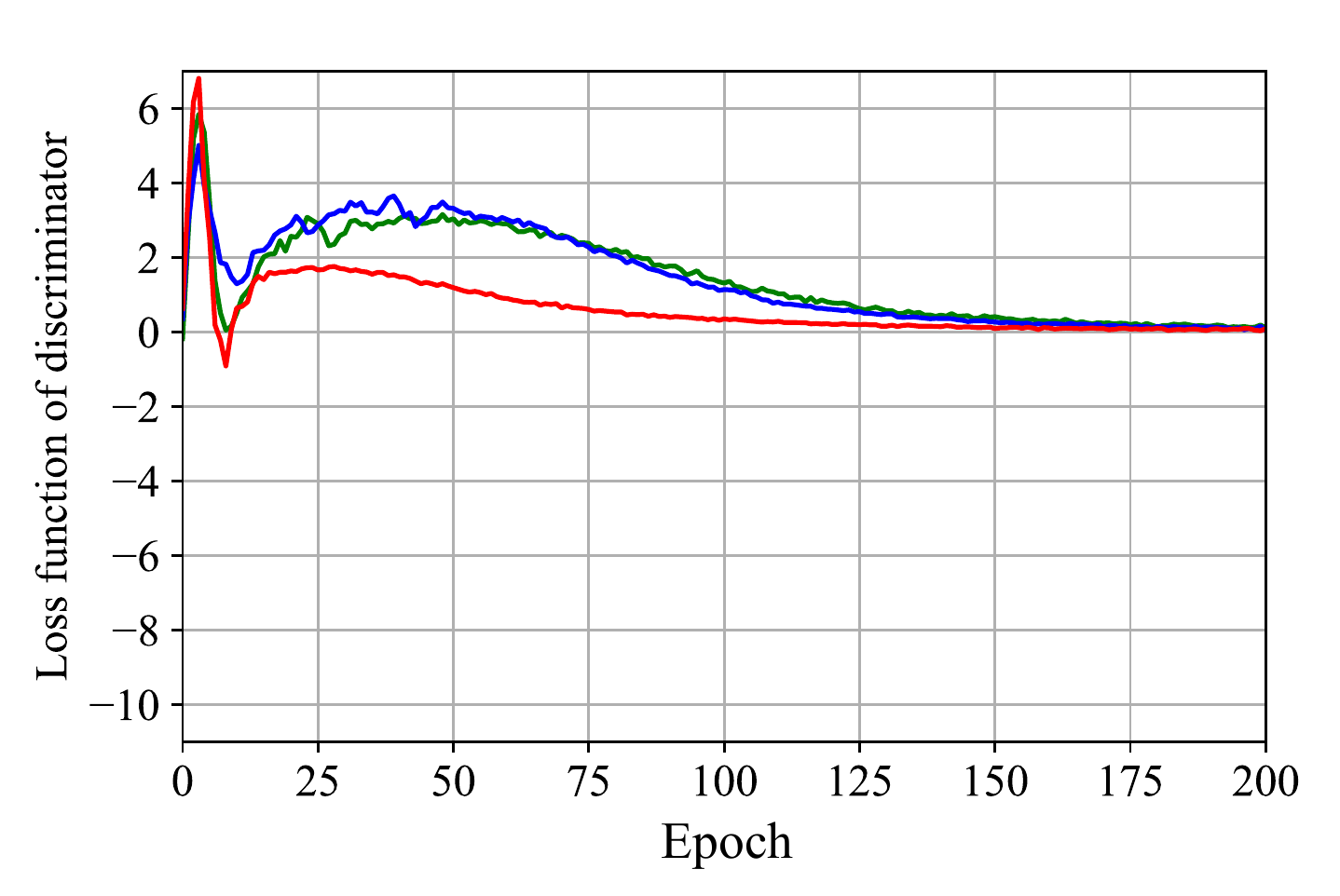}
        \caption{Discriminator}
    \end{subfigure}
    \caption{Quantitative assessment of the standard GAN, standard cGAN, and constrained cGAN, showing the training history of the loss functions associated with (a) the generator and (b) the discriminator. }
    \label{fig:M-lossfunctions}
\end{figure}

Furthermore, the evaluation metrics defined above are examined to provide a more quantitative assessment of the constrained versus the baseline GANs. The results are presented in Fig.\ref{fig:M-lossfunctions}. It can be seen in Figs.~\ref{fig:M-lossfunctions}a and~\ref{fig:M-lossfunctions}b that the loss functions of constrained cGAN converge faster, at approximately 125 epochs, while the loss functions of standard GANs and standard cGANs needed about 200 epochs to converge. Also note that the smaller noise level of the generator loss in Fig.~\ref{fig:M-lossfunctions}a suggests that the generated samples are less noisy. 
One epoch indicates that all training data has been used once, as each step of gradient descent in the training (weight optimization) uses only a small fraction (referred to as ``mini-batch'') of the training data. Larger epoch numbers correspond to longer training time. It can be seen that constrained GAN and cGAN have faster convergence.
This is because incorporating constraints generally restricts the solution space of the optimization problem onto a lower-dimensional manifold.

\begin{figure}[!htb]
    \centering
    \includegraphics[width=0.4\textwidth]{M-legend} \\
    \begin{subfigure}[b]{0.44\textwidth}
        \includegraphics[width=1\textwidth]{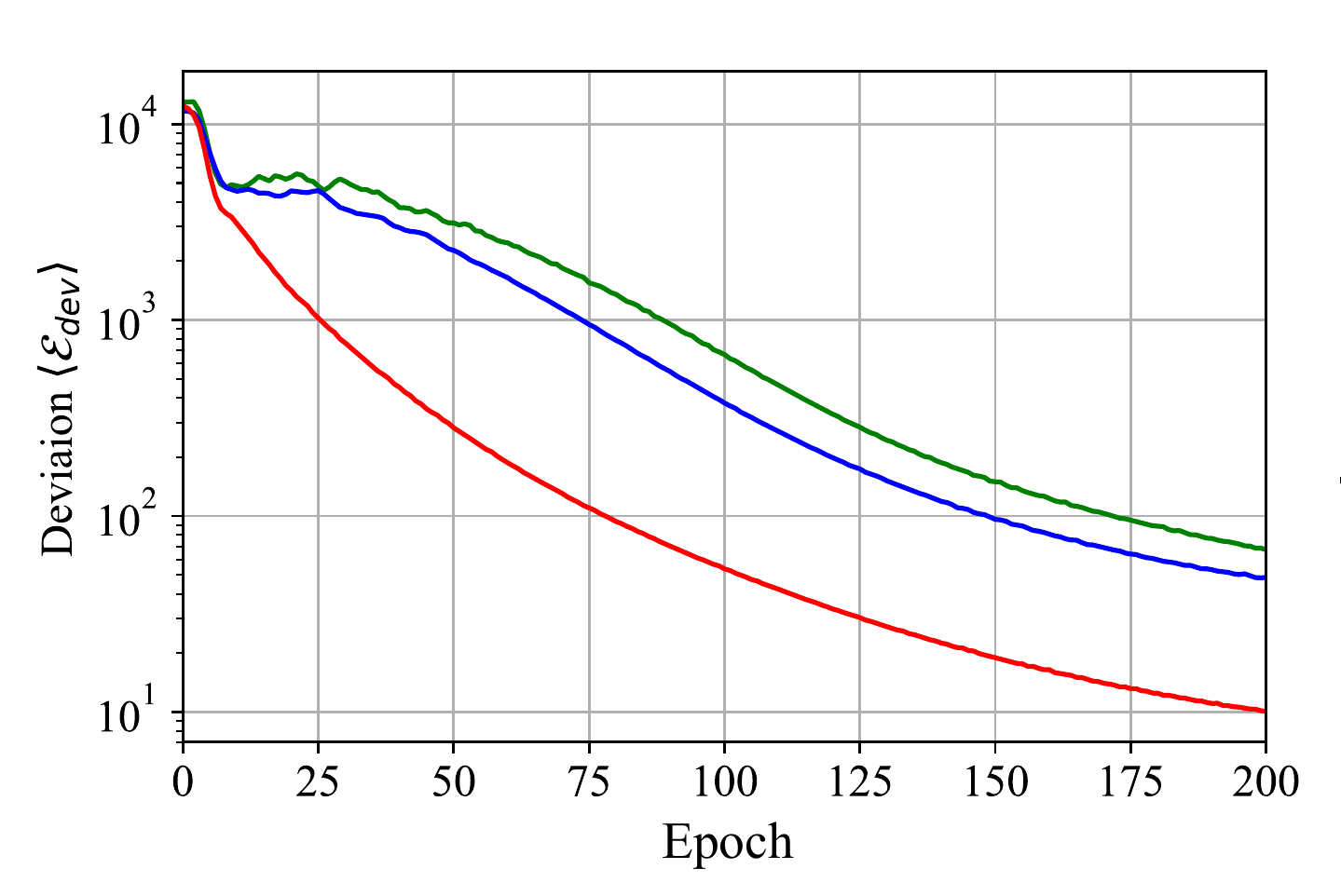}
        \caption{Deviations $\langle \mathcal{E}_\text{dev} \rangle$}
    \end{subfigure}
    \begin{subfigure}[b]{0.46\textwidth}
        \includegraphics[width=1\textwidth]{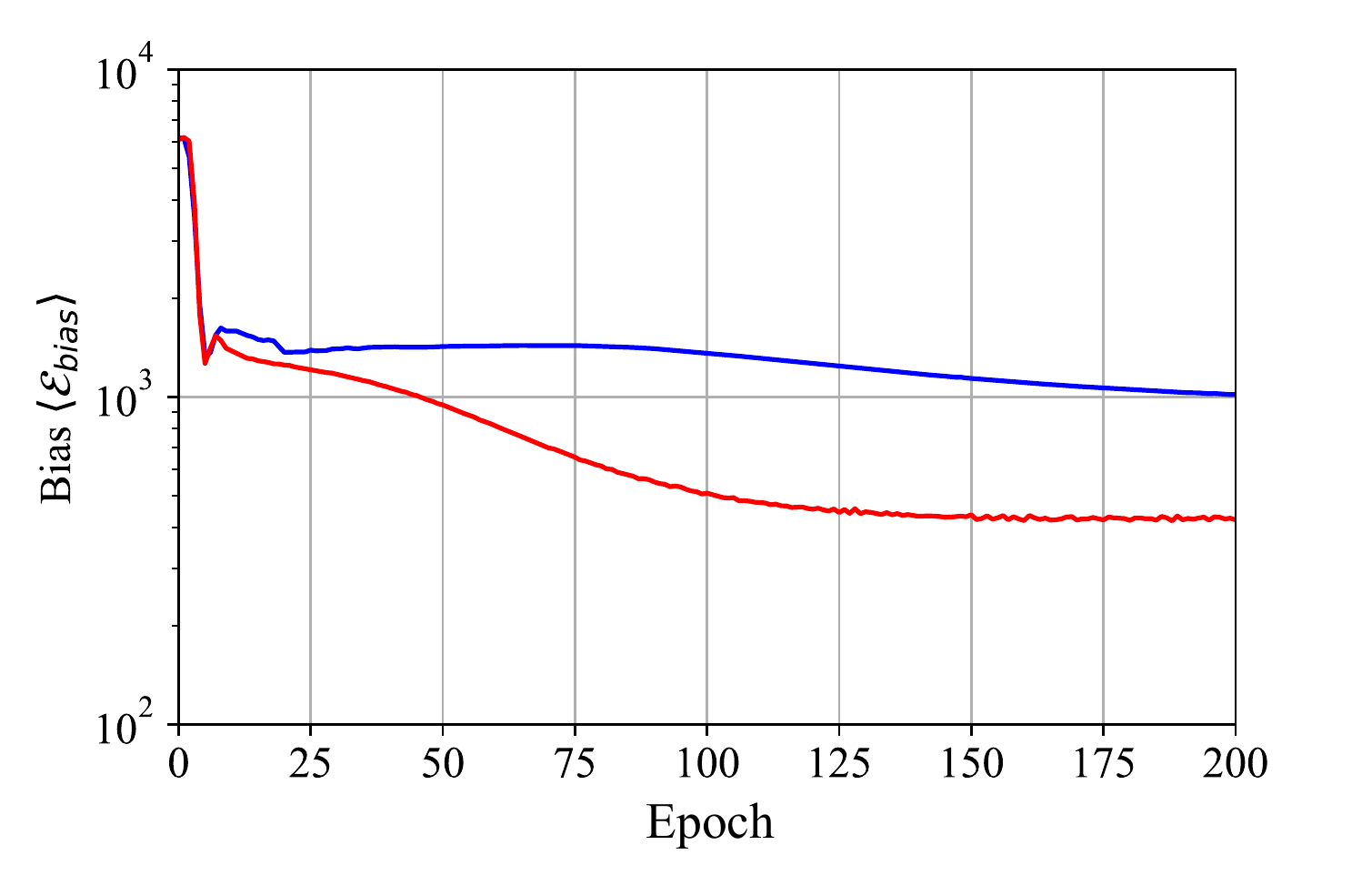}
        \caption{Bias $\langle \mathcal{E}_\text{bias} \rangle$}
    \end{subfigure}
    \caption{Quantitative assessment of the standard GAN, standard cGAN, and constrained cGAN, showing the training history (a) the deviations and (b) the biases as defined in Eqs.~\eqref{eq:M-deviation} and ~\eqref{eq:M-bias}, respectively. }
    \label{fig:M-dev-bias}
\end{figure}

However, the loss functions do not provide a direct quantification about the quality of generated samples.
In order to better compare the quality of generated samples, comparisons of deviation $\langle \mathcal{E}_\text{dev} \rangle$ and bias $\langle \mathcal{E}_\text{bias} \rangle$ are presented in Figs.~\ref{fig:M-dev-bias}. It can be seen in Fig.~\ref{fig:M-dev-bias}a that the deviation (a metric that quantifies the noise level of a generated sample) is slightly smaller for standard cGAN than standard GAN, and the deviation is about one order of magnitude smaller for the constrained cGAN. For the generated samples from cGANs, the biases as defined in Eq.~\ref{eq:M-bias} between the standard and constrained cGANs are compared in Fig.~\ref{fig:M-dev-bias}b. It can be seen that the biases of the generated sample from constrained cGAN are also noticeably smaller, indicating that the radius (the condition label) is better conformed to in the generated samples when the constraint is imposed.

\begin{figure}[htb]
    \centering
    \begin{subfigure}[b]{0.4\textwidth}
        \includegraphics[width=\textwidth]{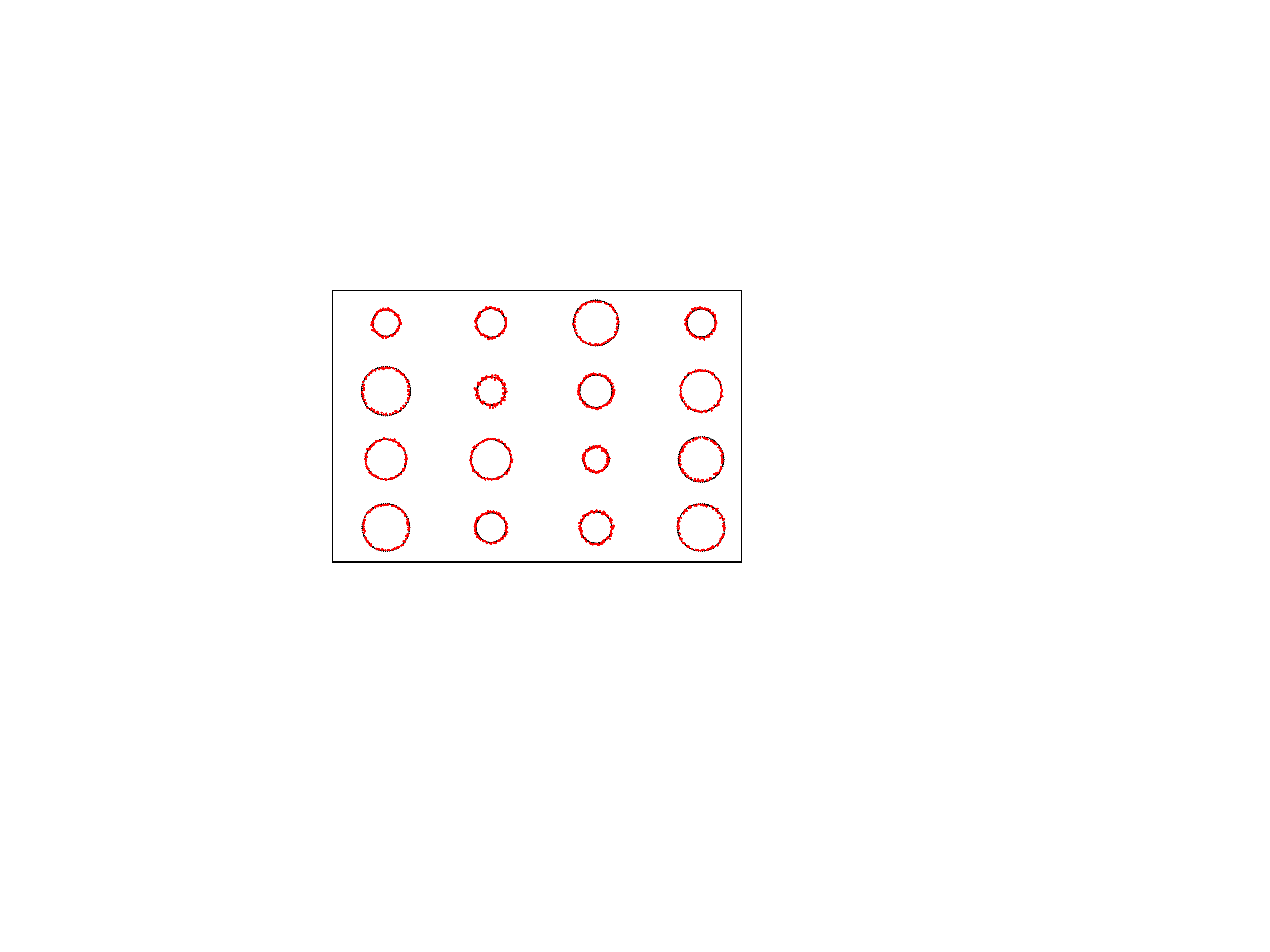}
        \caption{Constrained cGANs, $\varepsilon=0.1$}
    \end{subfigure}
    \hspace{3mm}
    \begin{subfigure}[b]{0.4\textwidth}   
        \includegraphics[width=\textwidth]{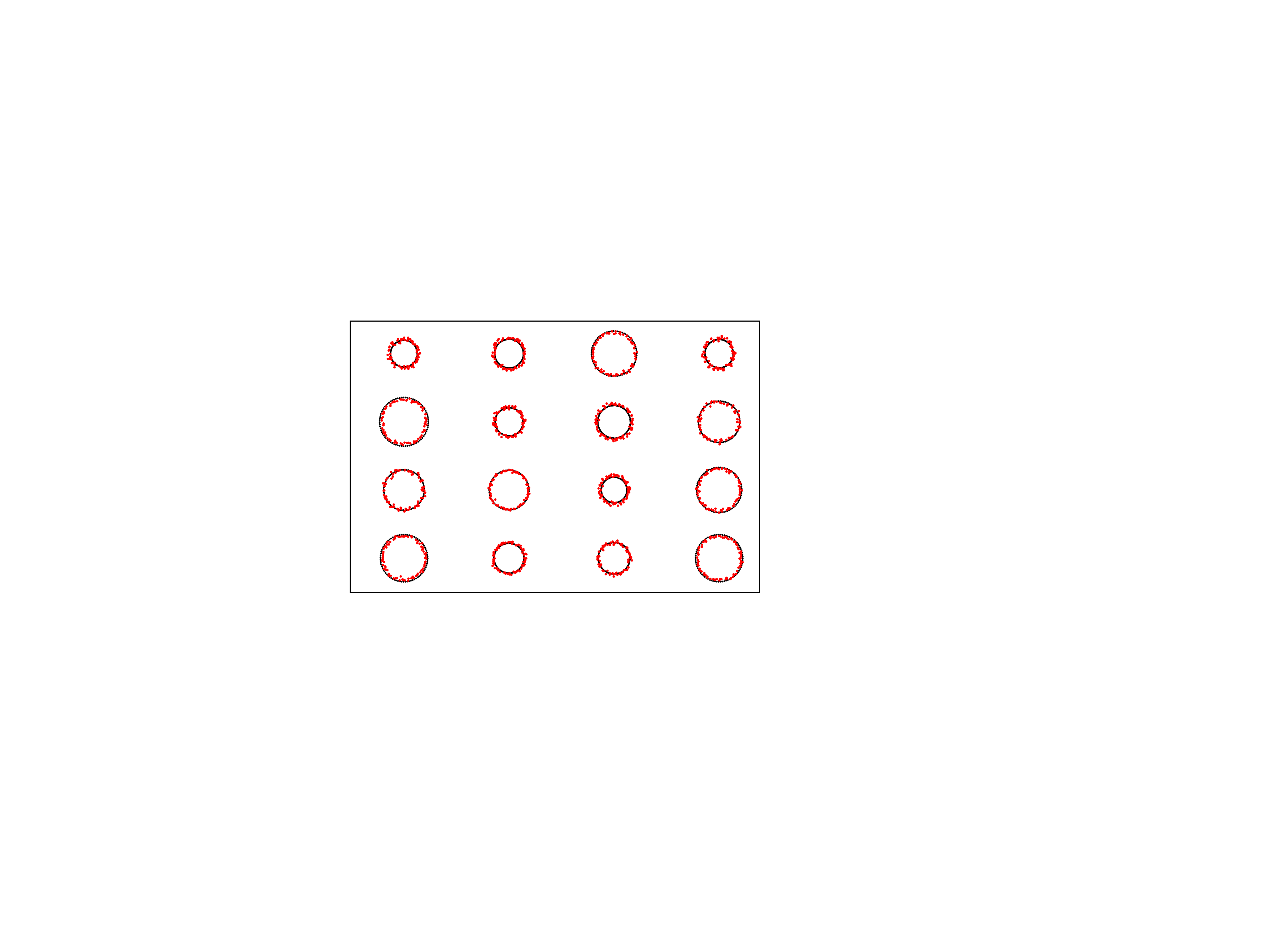}
        \caption{Constrained cGANs, $\varepsilon=0.2$}
    \end{subfigure}
    \caption{The generated samples (red/grey dots) by cGANs with different levels of imprecise constraints corresponding to $\varepsilon=0.1$ and $\varepsilon = 0.2$. The black/solid circles correspond to those with specified radius (condition labels).}
    \label{fig:M-generateddata-soft}
\end{figure}

The effects of imposing imprecise constraints (see Eq.~\eqref{eq:soft-circle} and Fig.~\ref{fig:circle-constraints}) are evaluated by setting positive tolerance parameters $\varepsilon$. We test two different tolerance values, $\varepsilon=0.1$ and $0.2$, to show that introducing imprecise constraint can improve the training performance of GANs. As can be seen in Fig.~\ref{fig:M-generateddata-soft}, the generated samples better represent circles than the standard cGAN results in Fig.~\ref{fig:Generated-data}b. The improvements are observed in both tolerance values. In addition, noises in the generated samples increase as the tolerance parameter $\varepsilon$ increase, which can be seen by comparing Fig.~\ref{fig:Generated-data}c (corresponding to $\varepsilon=0$), Fig.~\ref{fig:M-generateddata-soft}a ($\varepsilon=0.1$), and Fig.~\ref{fig:M-generateddata-soft}c ($\varepsilon=0.2$).
Such a trend is expected as the constraints are imposed less accurately with increasing tolerance $\varepsilon$. Indeed, the unconstrained, standard conditional GANs (cGANs, see Eq.~\eqref{eq:lossfunction-cGANs})~\cite{mirza2014conditional} (Fig.~\ref{fig:Generated-data}b) corresponds to $\varepsilon \to \infty$. On the other hand, the precise constraints (corresponding to $\varepsilon=0$) are often not available in practical physical modeling for reasons described in Section~\ref{sec:intro-enforce}. Here they are included merely as baseline for comparison.

\begin{figure}[htb]
    \centering
    \includegraphics[width=0.6\textwidth]{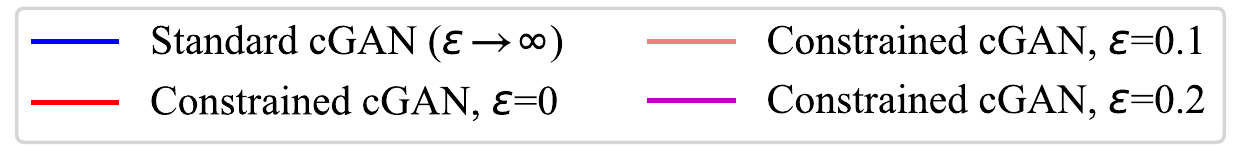} 
    \begin{subfigure}[b]{0.45\textwidth}
        \includegraphics[width=\textwidth]{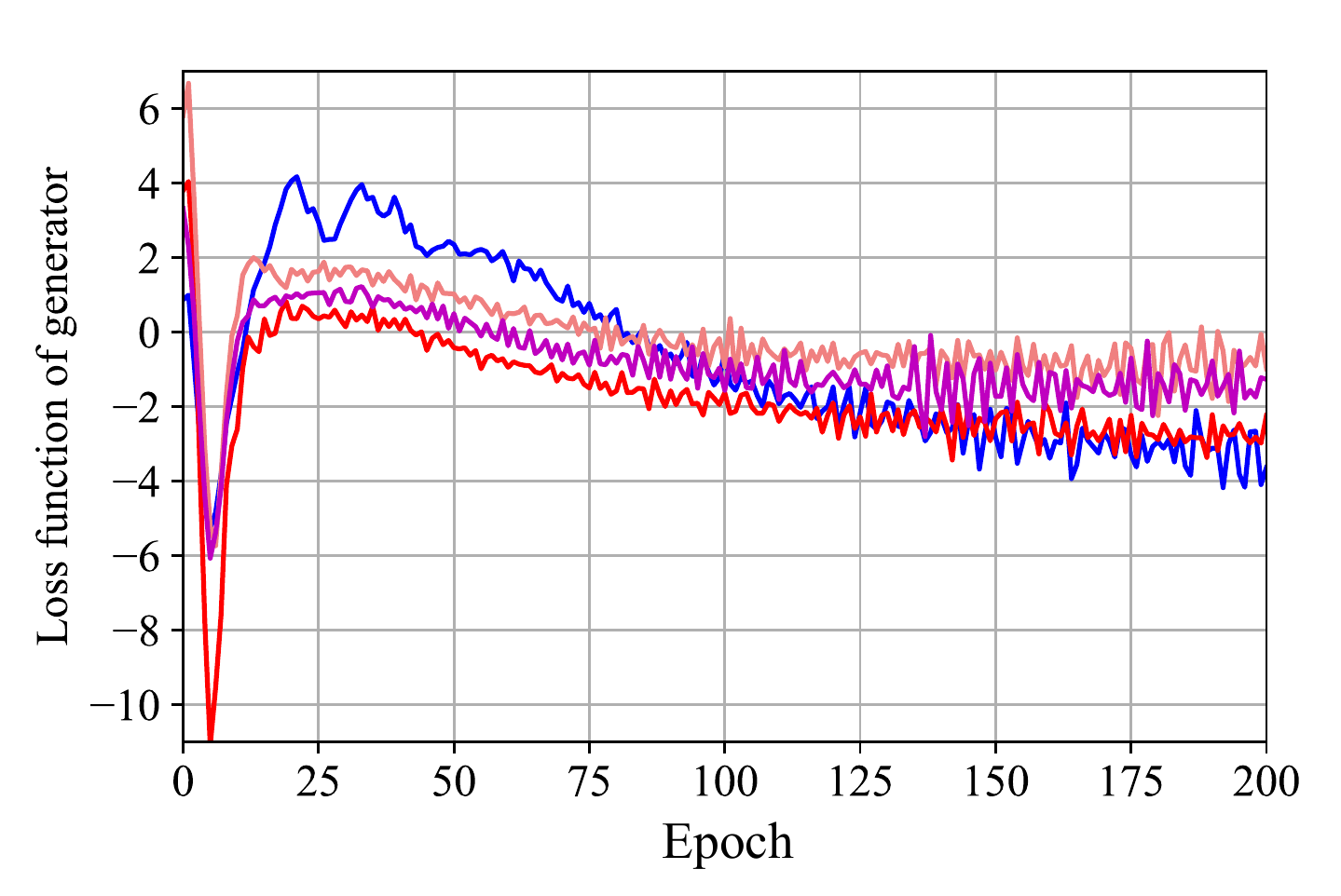}
        \caption{Loss function of generator}
    \end{subfigure}
    \hspace{3mm}
    \begin{subfigure}[b]{0.45\textwidth}   
        \includegraphics[width=\textwidth]{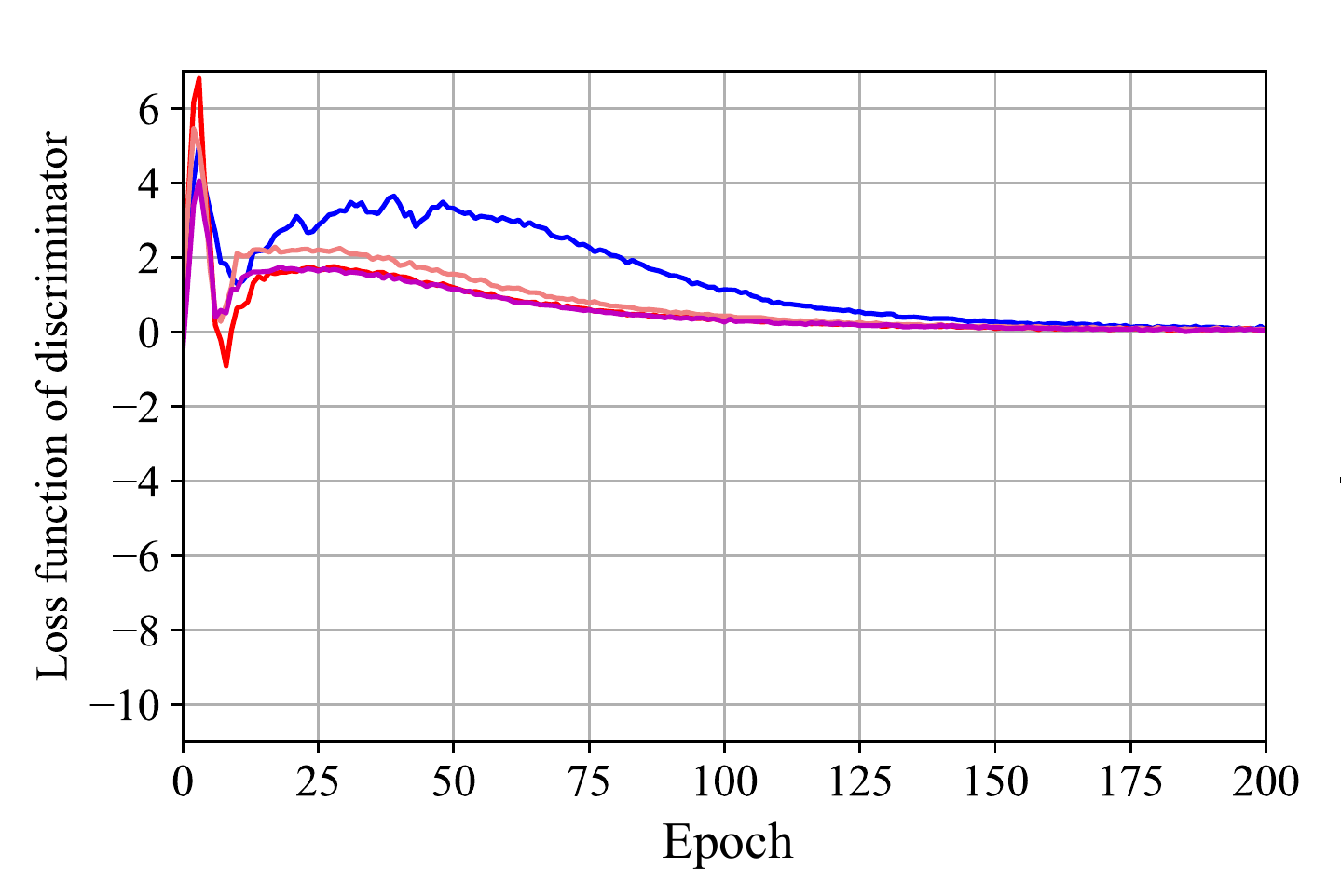}
        \caption{Loss function of discriminator}
    \end{subfigure}
    \hspace{2mm}
    \begin{subfigure}[b]{0.45\textwidth}
        \includegraphics[width=\textwidth]{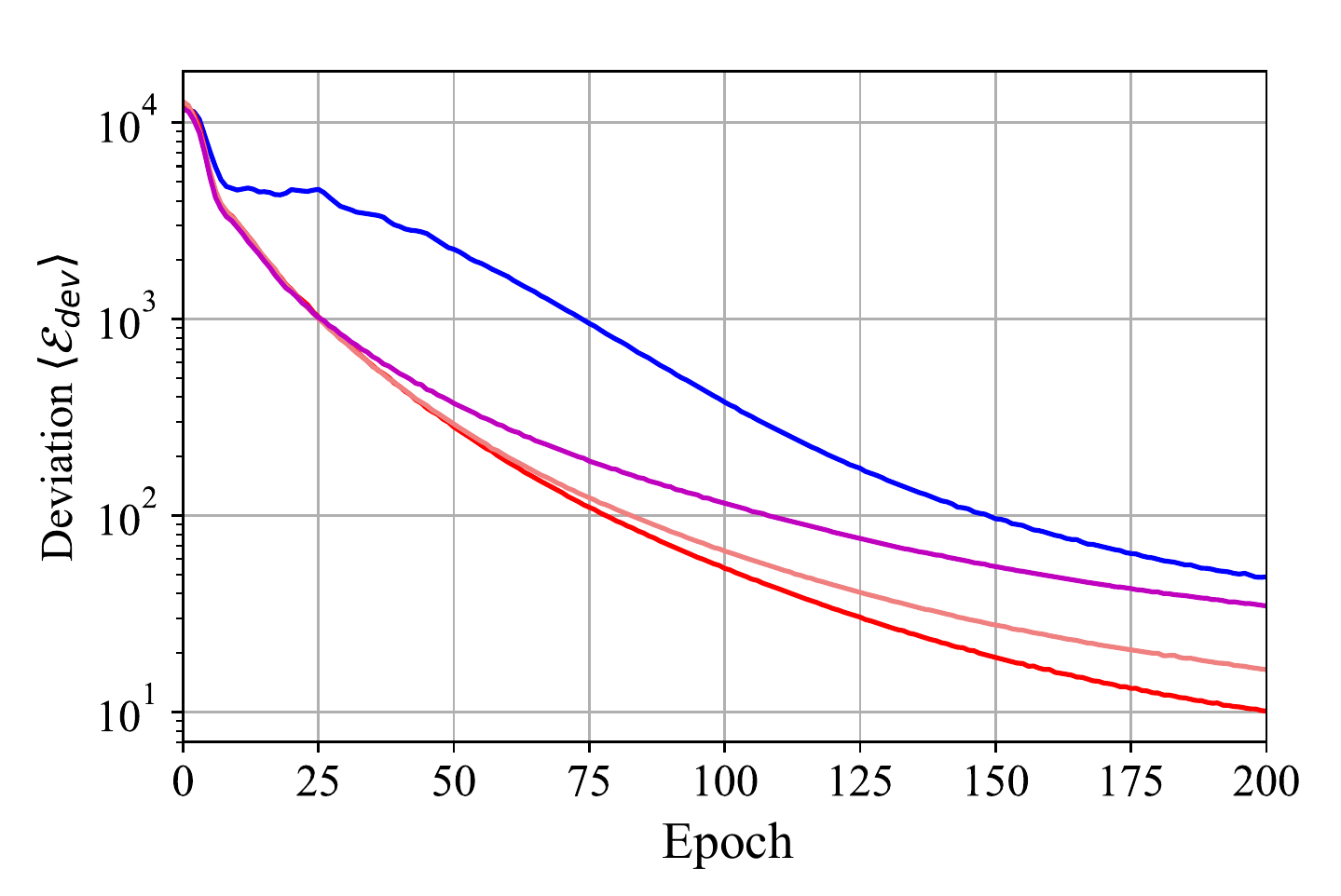}
        \caption{Deviations $\langle \mathcal{E}_\text{dev} \rangle$}
    \end{subfigure}
    \hspace{3mm}
    \begin{subfigure}[b]{0.47\textwidth}
        \includegraphics[width=\textwidth]{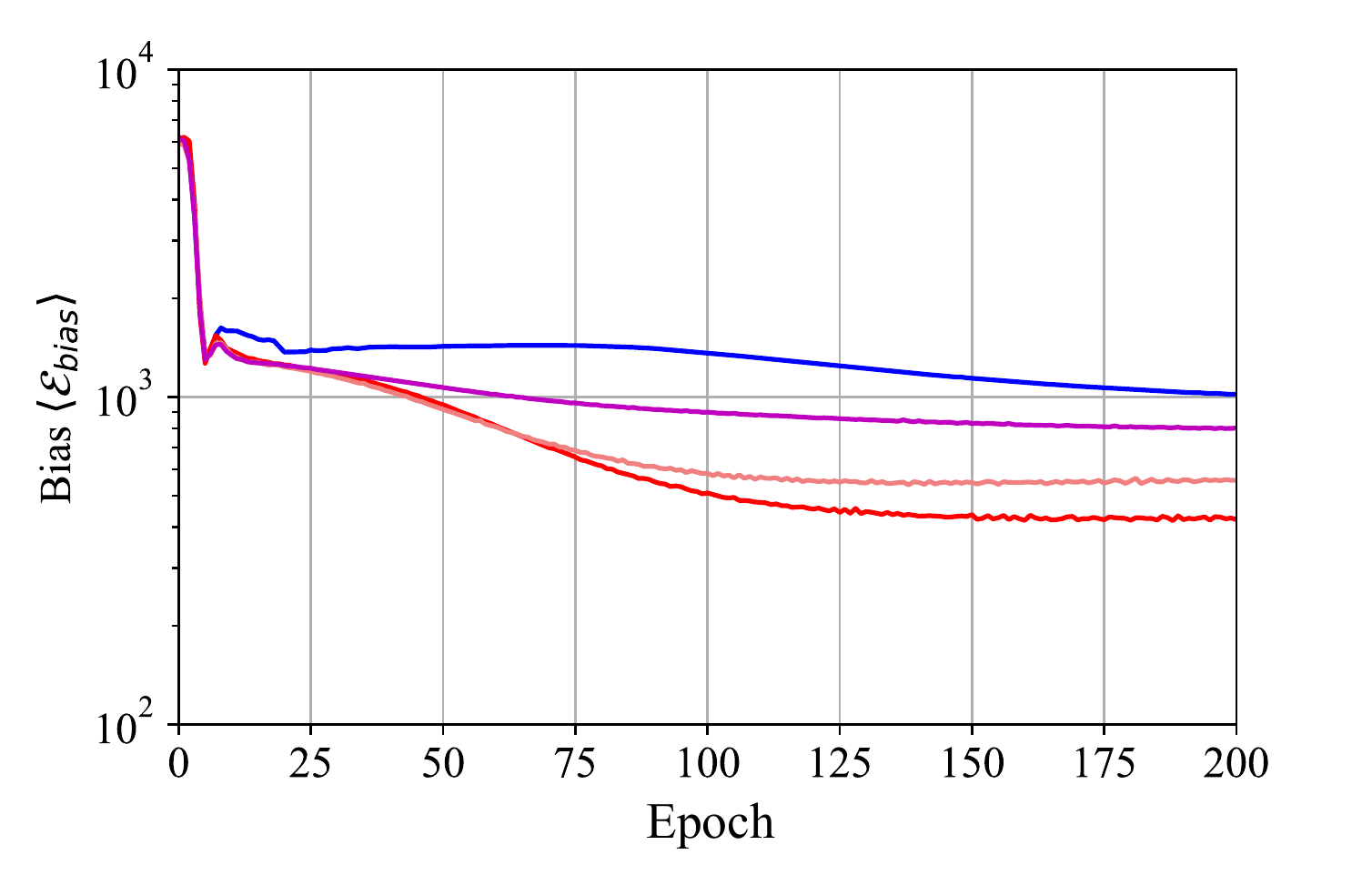}
        \caption{Bias $\langle \mathcal{E}_\text{bias} \rangle$}
    \end{subfigure}
    \caption{Effects of tolerance parameter $\varepsilon$ in imprecise constraints in the geometric constraint case. Metrics including (a) the loss functions of the generator, (2) the loss functions of the discriminator, along with (c) deviations and (d) biases at different tolerances for standard (unconstrained) cGANs~\cite{mirza2014conditional} and constrained cGANs.}
    \label{fig:M-lossfunction-soft}
\end{figure}

Quantitative metrics are further examined to illustrate the effect of adding imprecise constraints, with the results presented in Fig.~\ref{fig:M-lossfunction-soft}. It can be seen in Figs.~\ref{fig:M-lossfunction-soft}a and~\ref{fig:M-lossfunction-soft}b that the loss functions for both the generator and the discriminator demonstrate similar behaviors regardless of whether precise constraint ($\varepsilon=0$) or imprecise constraints ($\varepsilon^2 >0$) are imposed. This observation suggests that both types of constraints lead to similar improvements in terms of the convergence when training GANs. However, the quality of generated samples become more similar to the standard cGAN results as shown in Figs.~\ref{fig:M-lossfunction-soft}c and~\ref{fig:M-lossfunction-soft}d as the tolerance $\varepsilon$ increases.

\begin{figure}[!htb]
    \centering
    \includegraphics[width=0.7\textwidth]{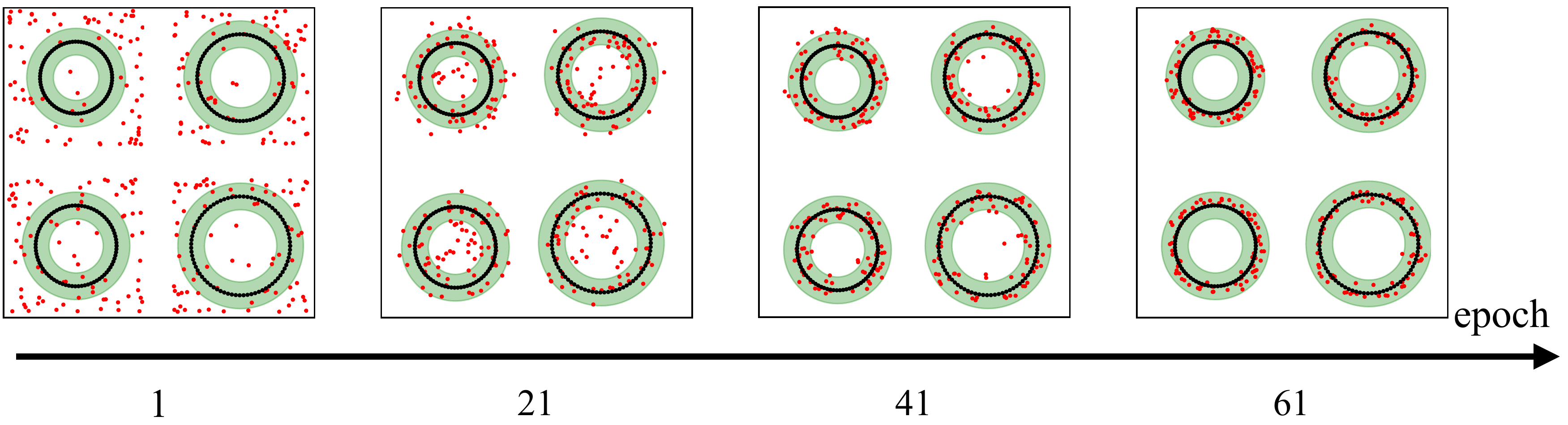}\\
    \vspace{1em}
    \begin{subfigure}[b]{0.35\textwidth}
    \centering
        \includegraphics[width=0.85\textwidth]{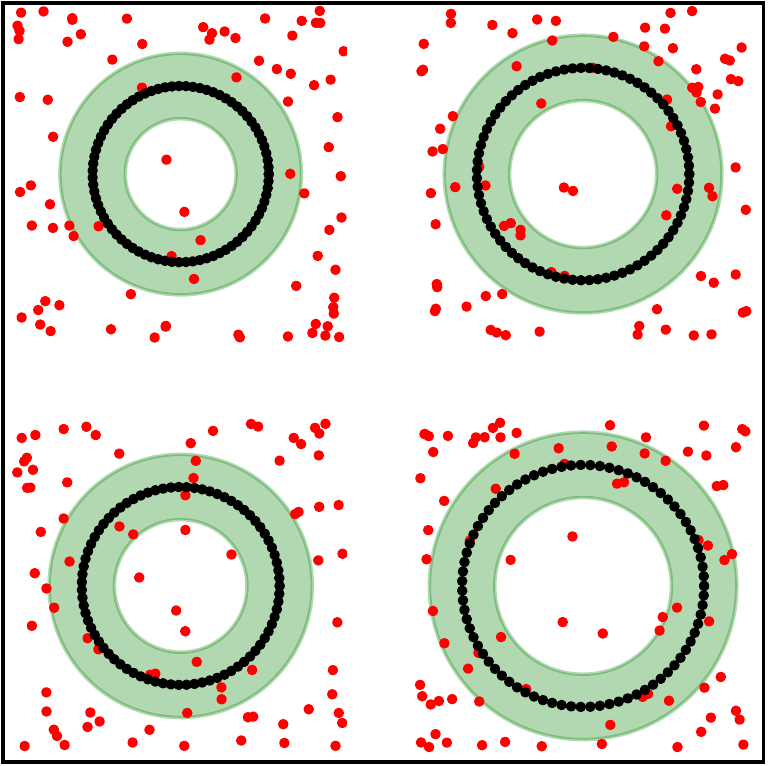}
        \caption{Epoch 1}
    \end{subfigure}
    \centering
    \begin{subfigure}[b]{0.35\textwidth}
    \centering
        \includegraphics[width=0.85\textwidth]{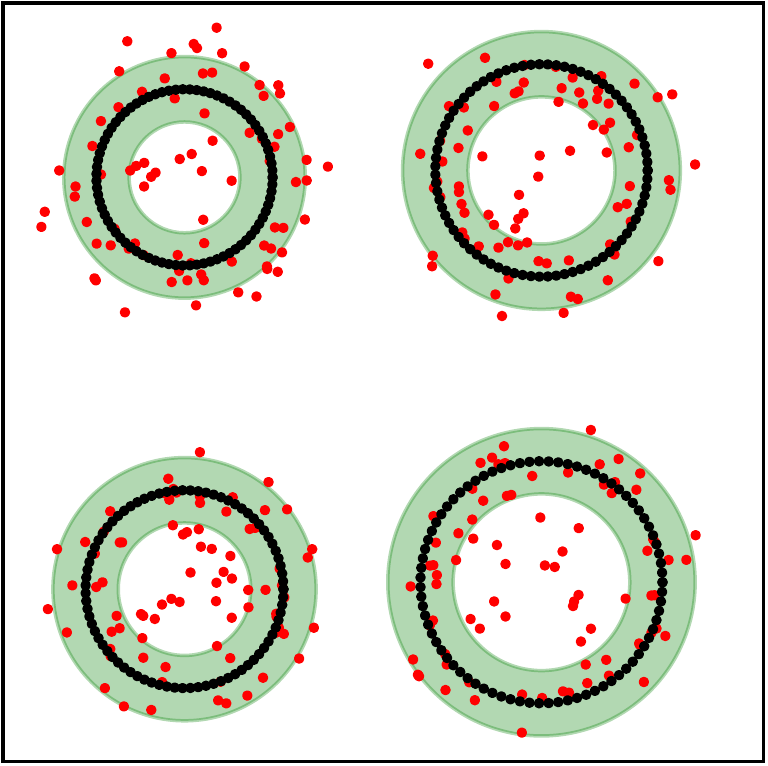}
        \caption{Epoch 21}
    \end{subfigure}
    
    \begin{subfigure}[b]{0.35\textwidth}
    \centering
        \includegraphics[width=0.85\textwidth]{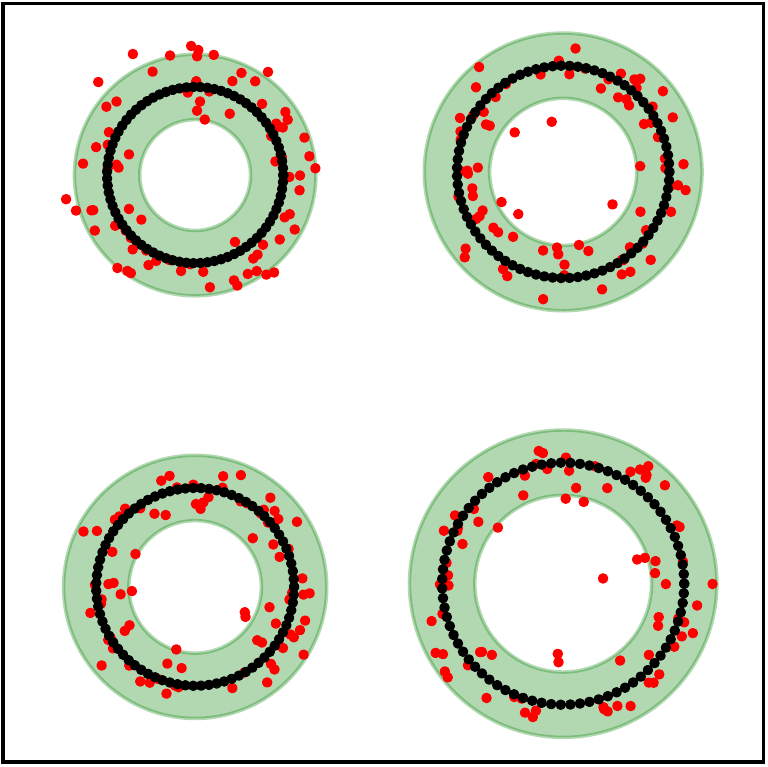}
        \caption{Epoch 41}
    \end{subfigure}
    \centering
    \begin{subfigure}[b]{0.35\textwidth}
    \centering
        \includegraphics[width=0.85\textwidth]{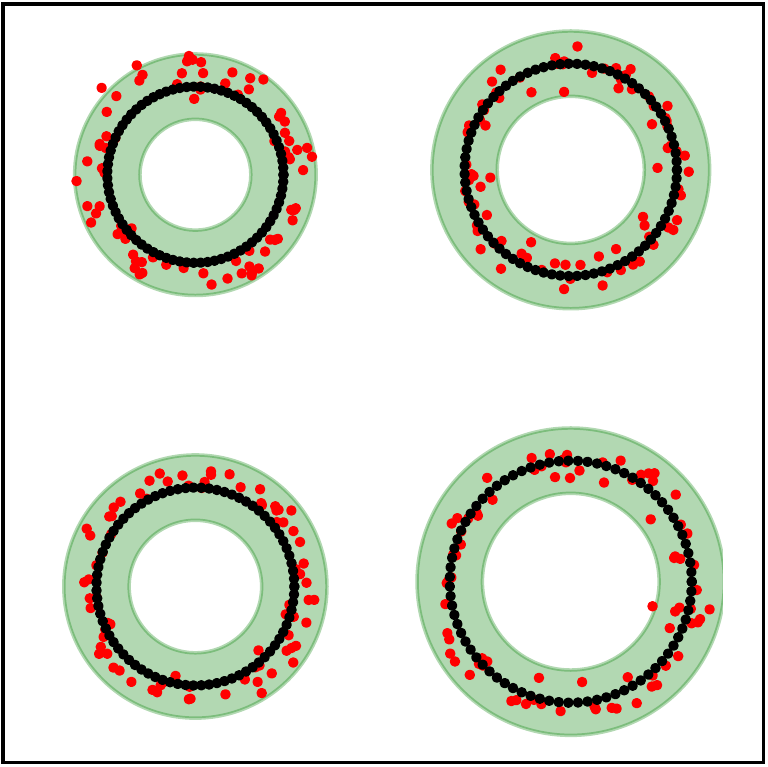}
        \caption{Epoch 61}
    \end{subfigure}
    
    \caption{The generated data at different epochs (1, 21, 41, and 61) during the initial stage of the training. Red/grey dots indicate the generated points by the constrained cGAN.  An imprecise constraint with a tolerance of $\varepsilon=0.4$ is imposed (indicated as shaded annulus). The black/dark dots (appearing as solid circles) indicate training samples, each of which consists of 100 points aligned on a circle of specified radius. }
    \label{fig:circles-epochs}
\end{figure}

We further examine the convergence of cGANs with imprecise constraints by analyzing samples during different stages (epochs) during the training, which are shown in Fig.~\ref{fig:circles-epochs}.  
Recall that the imprecise constraint is only active when generated points are located outside the shaded annulus (see Fig.~\ref{fig:circle-constraints}). The generated points scatter randomly at the beginning of training (at epoch 1) as shown in Fig.~\ref{fig:circles-epochs}a. During this stage of the training, the imprecise constraint term plays an important role as most points fall outside of the shaded annulus. The effect of the imprecise constraint term then gradually decay as the training proceeds, with more points falling within the shaded region. At epoch 21, the points start to concentrate toward the region within the annulus region (Fig.~\ref{fig:circles-epochs}b). The generated points further concentrate more within the shaded annulus region at epoch 41 (Fig.\ref{fig:circles-epochs}c). Eventually, almost all points concentrate within the annulus region, and most of them stick close to the circles of specified radii as indicated by black lines (Fig.\ref{fig:circles-epochs}d).  At this stage, most generated points fall within the shaded region and the imprecise constraint becomes negligible. As a result, the standard cGAN loss becomes dominant again and contributes further optimization of generated points within the shaded region.

\subsection{Differential constraint: generating potential flow fields}

In this test case, we aim to use GANs to generate velocity fields of fluid flows. This test case is motivated by current applications of GANs to generate inflow boundary conditions for large eddy simulations or hybrid LES/RANS simulations. 
In all such applications, being able to satisfy divergence-free condition in the generated velocity fields is the first metric. Further requirements such as statistical constraints are studied in a companion paper~\cite{wu2019enforcing}.

In this work we use samples of potential flow velocity fields as training data. The mass conservation (divergence-free condition for the velocity fields) is thus built into the data. In the constrained GANs, the imprecise constraint based on the divergence-free condition is embedded in the generator as described in Section~\ref{sec:method-implement}. The trained GANs, including both the standard GANs and the constrained GANs, are then evaluated by examining their generated samples in terms of (i) the degree to which they conform to the divergence-free condition and (ii) the smoothness of the generated velocity fields (because the training samples are smooth velocity fields).
While the divergence-free condition is embedded in the GANs as an imprecise constraint (see Eq.~\eqref{eq:soft-cons}), we emphasize that no smoothness constraints are  explicitly imposed in the network. Rather, it is expected that GANs would learn it from the data, as all the generated flow fields are smooth. This study also aim to investigate whether the imprecise divergence-free constraint can improve the smoothness of the generated flow fields.

\subsubsection{Problem description}
Two-dimensional velocity field training samples are generated by using the following analytic function~\cite{currie2002fundamental}:
\begin{equation}
    F(z) = c e^{-i\alpha z} + \frac{m}{2\pi}\log (z - z_0)
    \label{eq:phi}
\end{equation}
where $z = x+iy$. The complex potential in Eq.~\eqref{eq:phi} above is a superposition of a uniform flow potential (first term on the RHS) and a source flow potential (the second term on the RHS). The complex potential~$F$ is parameterized by four parameters: $c$ determines the magnitude of the uniform flow velocity, $\alpha$
determines the direction of the uniform flow, $m$ determines the strength of the source, and $z_0$ dictates the location of source. Without loss of generality, $z_0$ can be specified as the origin $(0,0)$, so $z_0$ is no longer a free parameter hereafter. By sampling the parameter vector $\{c, \alpha, m\}$, we can obtain various velocity fields. Specifically, denoting the real part of $F$ as $\phi$ (referred to as the \emph{velocity potential}), the velocity fields can be obtained by taking the gradient of $\phi$, i.e., $\bm{v} = \nabla \phi$.

\begin{figure}[!htb]
    \centering
    \includegraphics[width=0.45\textwidth]{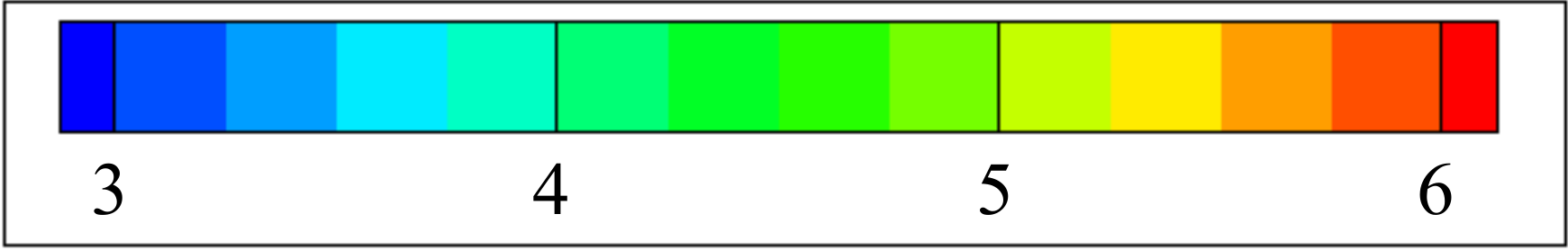}
     \includegraphics[width=0.6\textwidth]{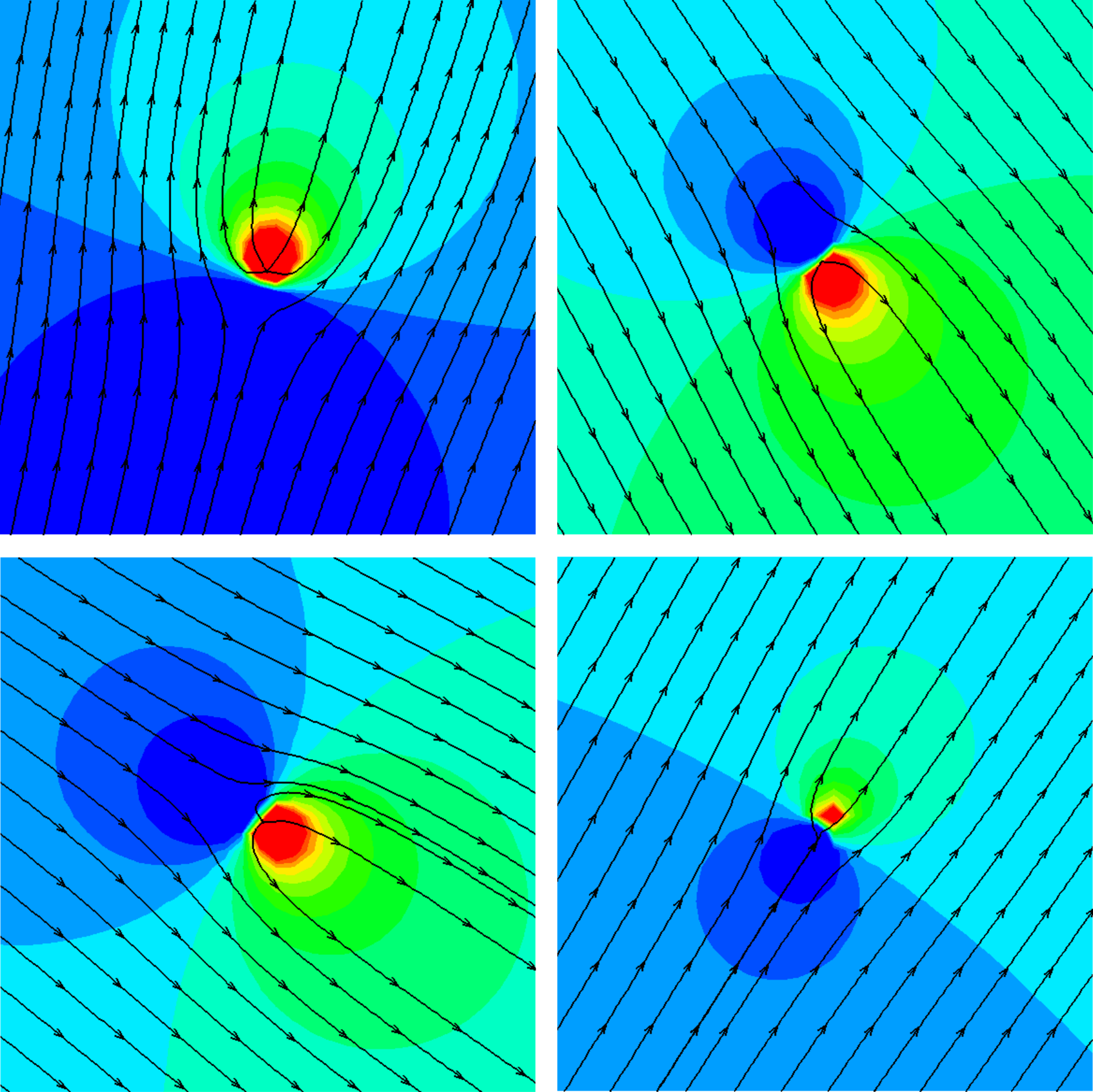}
    \caption{
    Some randomly selected training samples in the differential constraint case, showing velocity magnitude contours and streamlines corresponding to the flow velocity fields. The velocities are computed from the potential in Eq.~\eqref{eq:phi} with varying parameters. Each sample consists of a two-dimensional velocity field on a mesh of $32\times 32$ grid points.}
    \label{fig:N-training}
\end{figure}

The parameters $c$, $\alpha$ and $m$ are sampled from three independent Gaussian distributions~$N(\mu, \sigma^2)$ as follows:
\begin{equation}
    c\sim N(4, \, 0.4), \qquad 
    \alpha\sim N(0, \, \frac{\pi}{4}) , \qquad \text{and}  \qquad
    m \sim N(1, \, 0.2) ,
\end{equation}
with $\mu$ and $\sigma^2$ denoting the mean and  variance of the Gaussian distribution.
Each parameter combination $\{c, \alpha, m\}$ corresponds to a unique velocity field. 
The generated velocity are represented on a grid of $32 \time 32$ mesh points in the domain of $[-0.5, 0.5] \times [-0.5, 0.5]$ in the horizontal ($x$-)  and vertical ($y$-) directions. The origin of the presented domain is co-located with the source as specified above, i.e., $z_0 = (0, 0)$. The discretized velocity fields on the mesh are used as input to the GANs for training. We emphasize that the GANs are not aware of the parameterization of the velocity. All they see are the data in its high-dimensional representation, i.e., velocity  field on the mesh. We used 20000 velocity fields as training samples of the GANs, from which four random yet representative samples are shown in Fig.~\ref{fig:N-training}.

The imprecise constraint ensures the velocity divergence-free condition up to an accuracy $\varepsilon$, which is enforced via the following penalty term based on Eq.~\eqref{eq:soft-cons}:
\begin{equation}
   C_\text{phys} = E_{z\sim p_\text{z}(z)}[\max\left( \|\nabla \cdot \bm{v}\|_\text{F}^2, \; \varepsilon^2 \right)]
   \label{eq:loss-div}
\end{equation}
where $\varepsilon$ controls the tolerance of the divergence-free condition, $\|\cdot\|_\text{F}$ denotes the Frobenius norm of the velocity divergence field in the entire computational domain. The divergences $\nabla \cdot \bm{v}$ of the generated flow fields are computed via finite difference on the underlying uniform Cartesian grid, with central and one-sided schemes used for interior and boundary points, respectively.
Other setups and parameters used for the GAN are presented in Table~\ref{tab:GAN-settings-p}.
Convolutional and deconvolutional neural networks with four layers are used for the discriminator and generator, respectively. The detailed architectures of the two networks are presented in Table~\ref{tab:nn-details}.

\begin{table}[htbp]
\caption{Detailed settings and parameters of the GANs used in generating potential flow fields.}
\centering
\begin{tabular}{P{6cm} P{6cm}}
\hline
Loss function for baseline GAN & WGAN-GP \\
Dimension of latent space & $Z \in \mathbb{R}^{100}$ \\
Activation functions & LeakyReLU$^{\text{(a)}}$\\
Epochs & 200\\
Learning rate ($\eta$) &  0.005, 0.0005 \\
\hline
\end{tabular}
\begin{flushleft}
\small
Notes: (a) The slope of the LeakyReLU activation function is 0.2 for negative inputs.\\
\end{flushleft}
\label{tab:GAN-settings-p}
\end{table}

\begin{table}[htbp]
\caption{Detailed network architecture for the generator and discriminator. Acronyms used: C, number of channels; KS, kernel size; BN, batch normalization.}
\centering
\begin{tabular}{P{4cm} P{4cm} P{4cm}}
\hline
 & Generator & Discriminator \\
\hline \hline
Layer 1 & Deconvolutional, C:64, KS: $4\times4$, BN, LeakyReLU & Convolutional,  C:64, KS: $5\times5$, BN, LeakyReLU  \\
\hline
Layer 2 & Deconvolutional,  C:64, KS: $5\times5$, BN, LeakyReLU & Convolutional, C:64, KS:$5\times5$, BN, LeakyReLU  \\
\hline
Layer 3 & Deconvolutional, C:64, KS:$5\times5$, BN, LeakyReLU & Convolutional, C:64, KS:$5\times5$, BN, LeakyReLU  \\
\hline
Layer 4 & Deconvolutional, \hspace{1em} C:3, KS: $4\times4$  & Convolutional, \hspace{2em} C:1, KS: $5\times5$  \\
\hline
\end{tabular}
\label{tab:nn-details}
\end{table}

As an unsupervised learning method, GANs aim to generate samples that conform to the distribution of the training data. Therefore, it is not possible to assess the quality of the trained network by comparing the discrepancy of each generated flow field sample to the corresponding ``truth", since the latter does not exist. Rather, we need to devise metrics to assess how well the generated samples conform to the data distribution. To this end, we propose the following two metrics:
\begin{enumerate}[(a)]
    \item the degree to which they conform to these constraints, which is quantified by the norm of the velocity divergence in the field, i.e.,
    \begin{equation}
        \mathcal{E}_\text{div} = \| \nabla \cdot \bm{v} \|_\text{F} .
          \label{eq:ediv}
    \end{equation}
    A scalar metric is obtained above by taking the Frobenius norm of the velocity divergence in the entire domain, i.e., by summing the divergence of all cells. As in the geometric constraint case, we compute the sample mean $\langle \mathcal{E}_\text{div} \rangle$ by averaging that of all generates samples, where $\langle \cdot \rangle$ indicates ensemble averaging.  Field contours are also presented to show the spatial distribution of the divergence.
    \item the non-smoothness of a generated velocity field as measured by the gradient of the velocity magnitude:
    \begin{equation}
        \mathcal{E}_\text{grad} = \| \nabla v  \|_\text{F} ,
        \label{eq:egrad}
    \end{equation}
where $v = |\bm{v}|$ is the velocity magnitude, and the Frobenius is similarly defined as above. Similarly, the mean $\langle \mathcal{E}_\text{grad} \rangle$ of the above quantity is used to assess the quality of the generated samples.
\end{enumerate}

\subsubsection{Results and discussion}
\label{sec:res-discuss}

\begin{figure}
    \centering
    \includegraphics[width=0.32\textwidth]{case-cbar-velocity}
    
    \begin{subfigure}[b]{0.37\textwidth}
    \centering
        \includegraphics[width=1\textwidth]{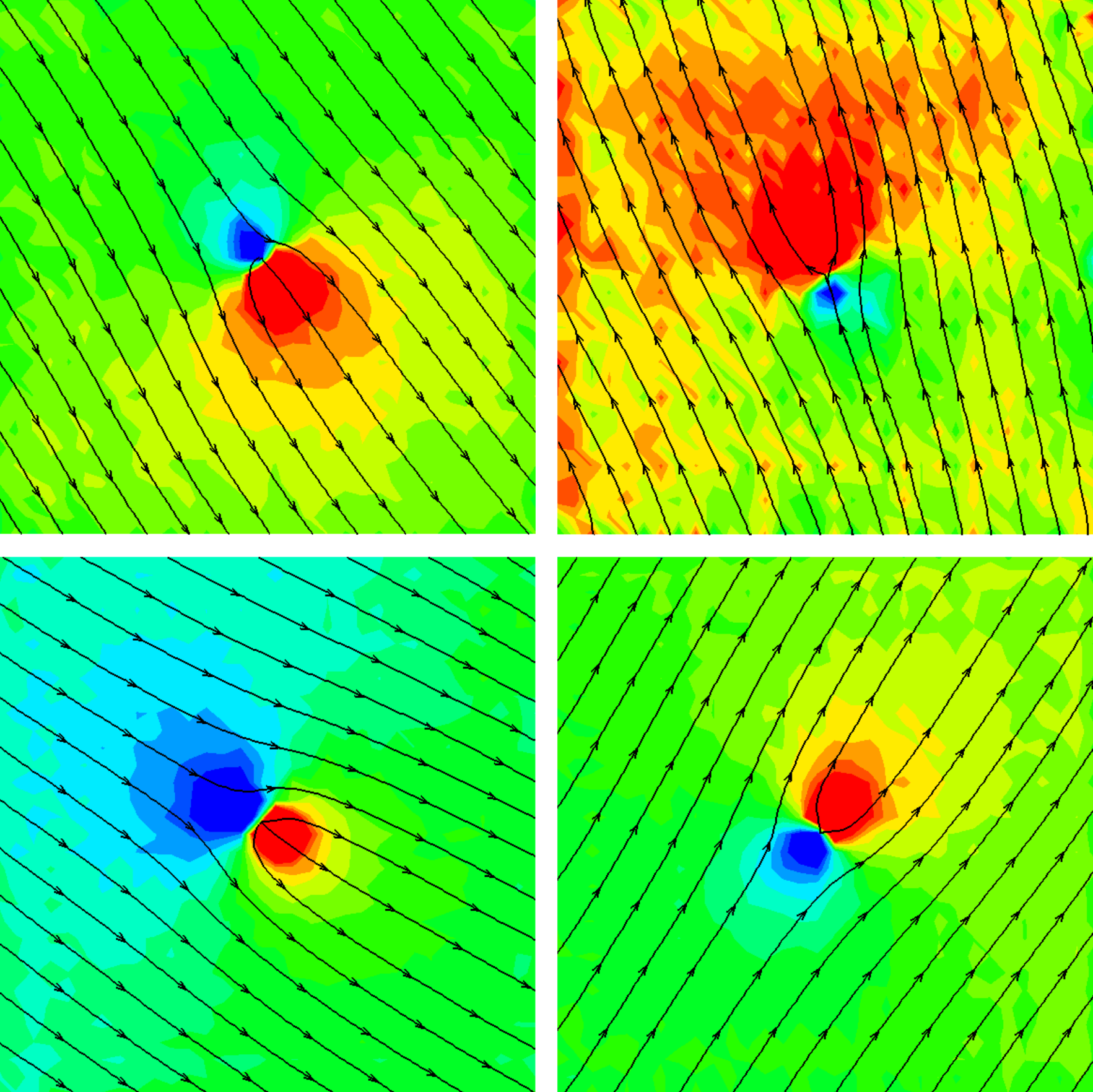}
        \caption{Standard GANs, epoch 40}
    \end{subfigure}
    \centering
    \begin{subfigure}[b]{0.37\textwidth}
    \centering
        \includegraphics[width=1\textwidth]{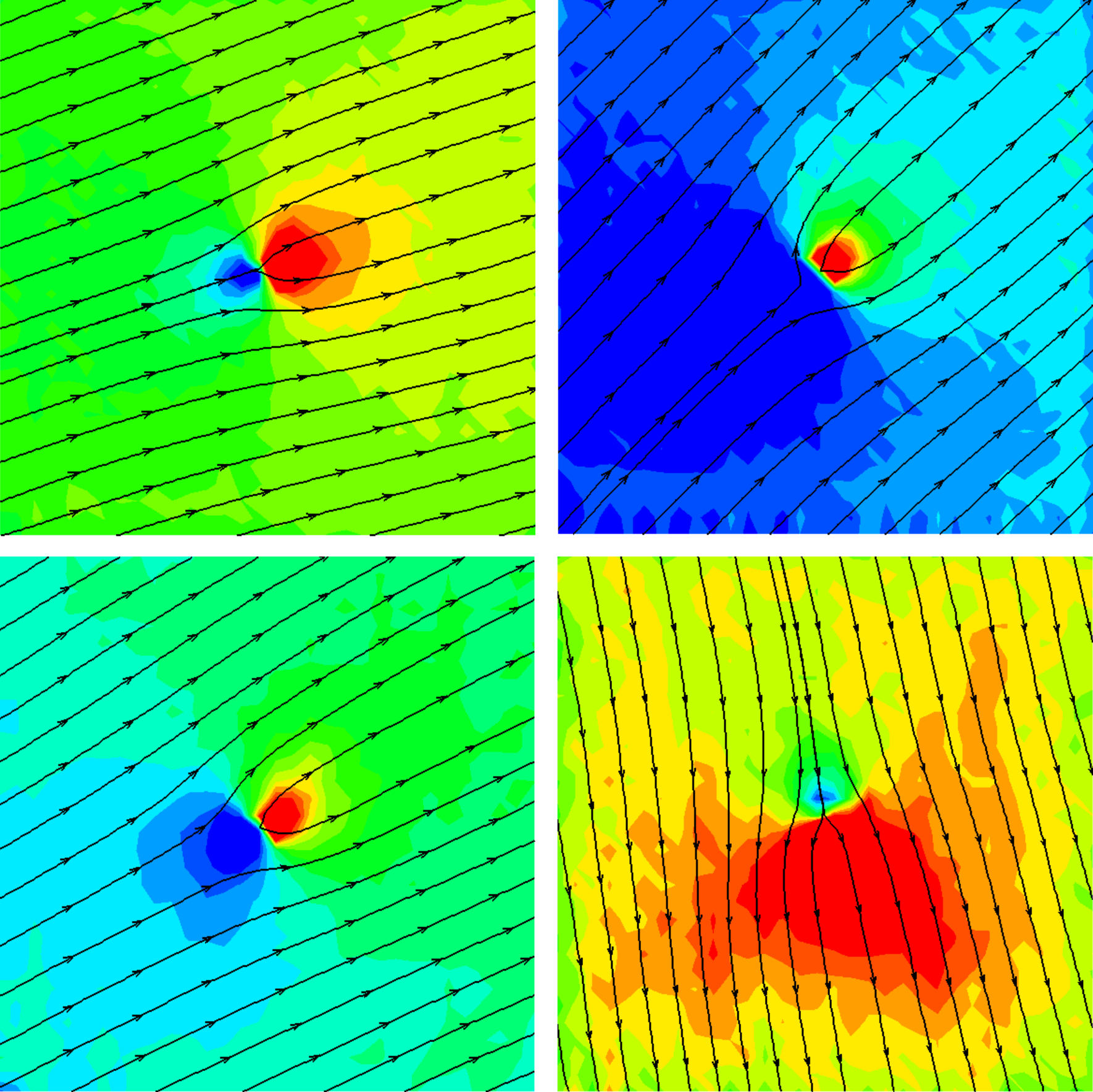}
        \caption{Constrained GANs, epoch 40}
    \end{subfigure}
    
    \begin{subfigure}[b]{0.37\textwidth}
    \centering
        \includegraphics[width=1\textwidth]{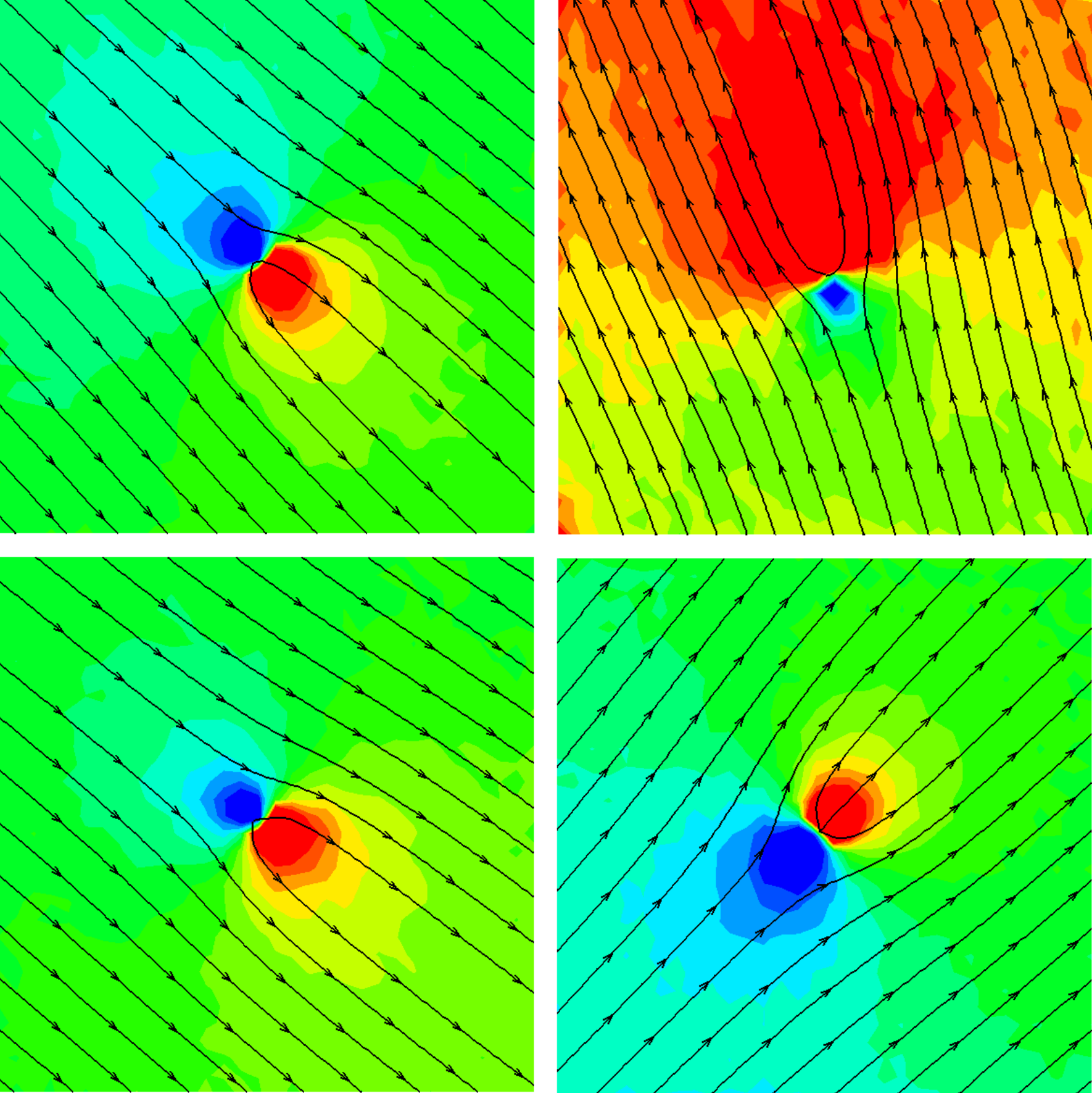}
        \caption{Standard GANs, epoch 80}
    \end{subfigure}
    \centering
    \begin{subfigure}[b]{0.37\textwidth}
    \centering
        \includegraphics[width=1\textwidth]{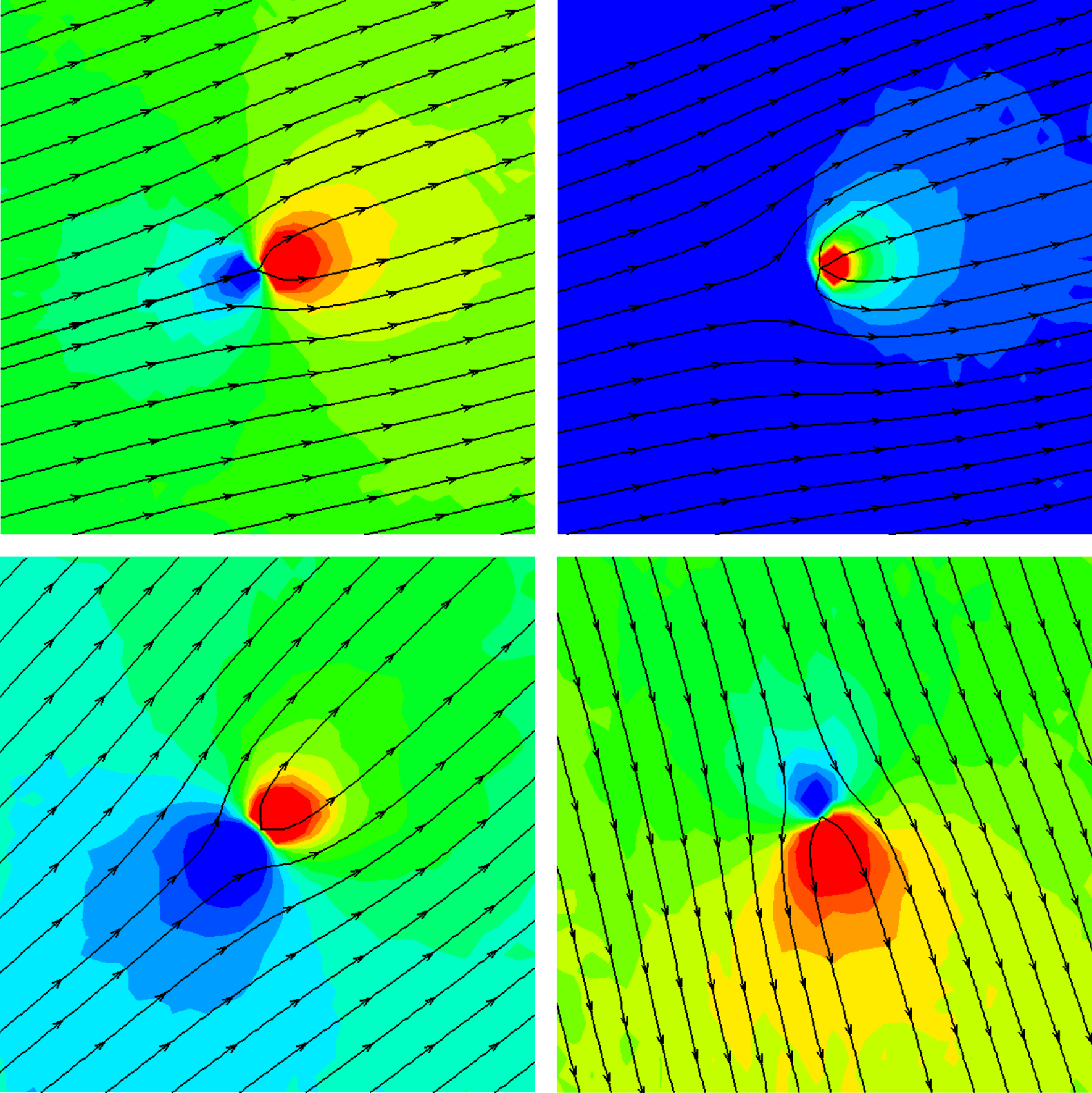}
        \caption{Constrained GANs, epoch 80}
    \end{subfigure}
    
    \begin{subfigure}[b]{0.37\textwidth}
    \centering
        \includegraphics[width=1\textwidth]{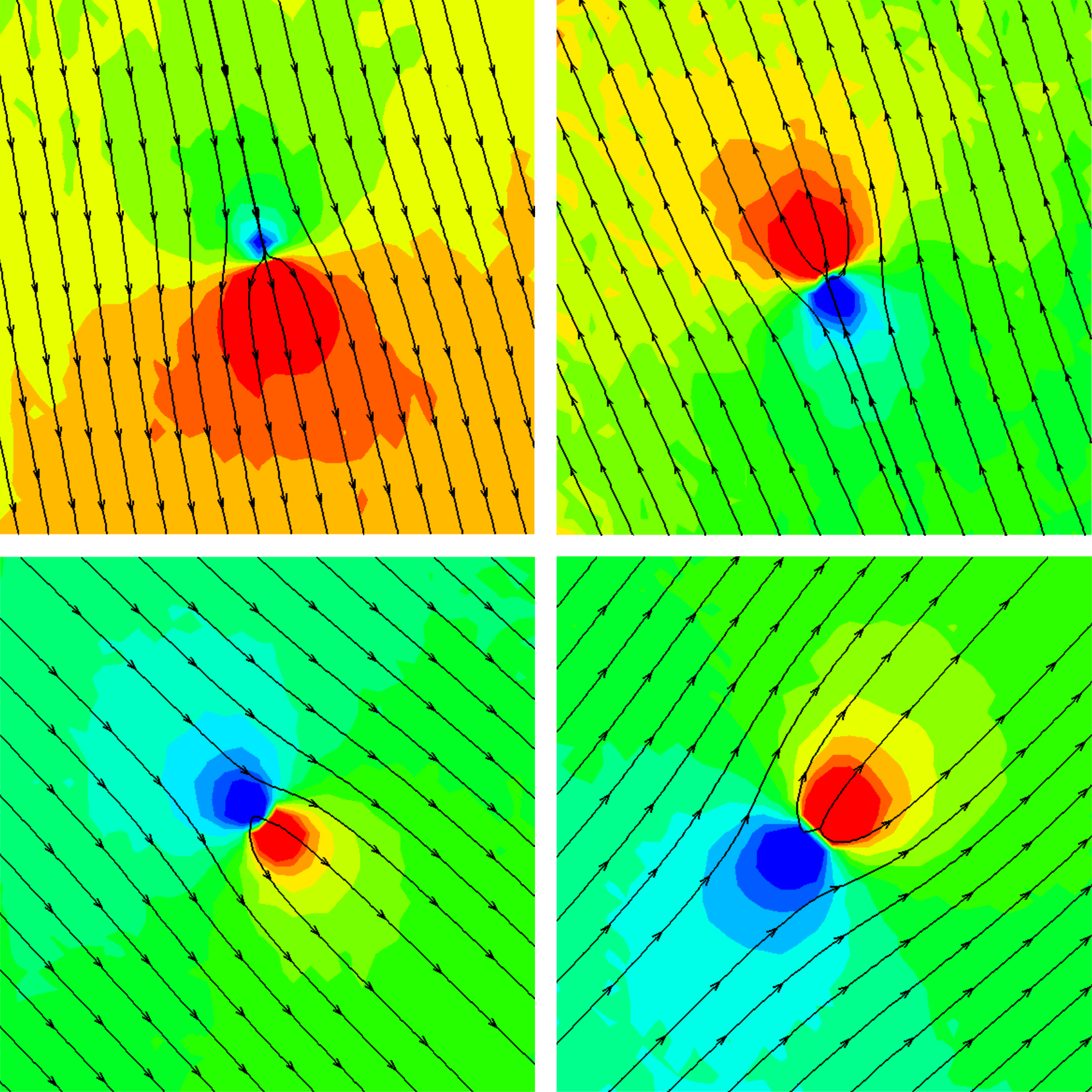}
        \caption{Standard GANs, epoch 120}
    \end{subfigure}
    \centering
    \begin{subfigure}[b]{0.37\textwidth}
    \centering
        \includegraphics[width=1\textwidth]{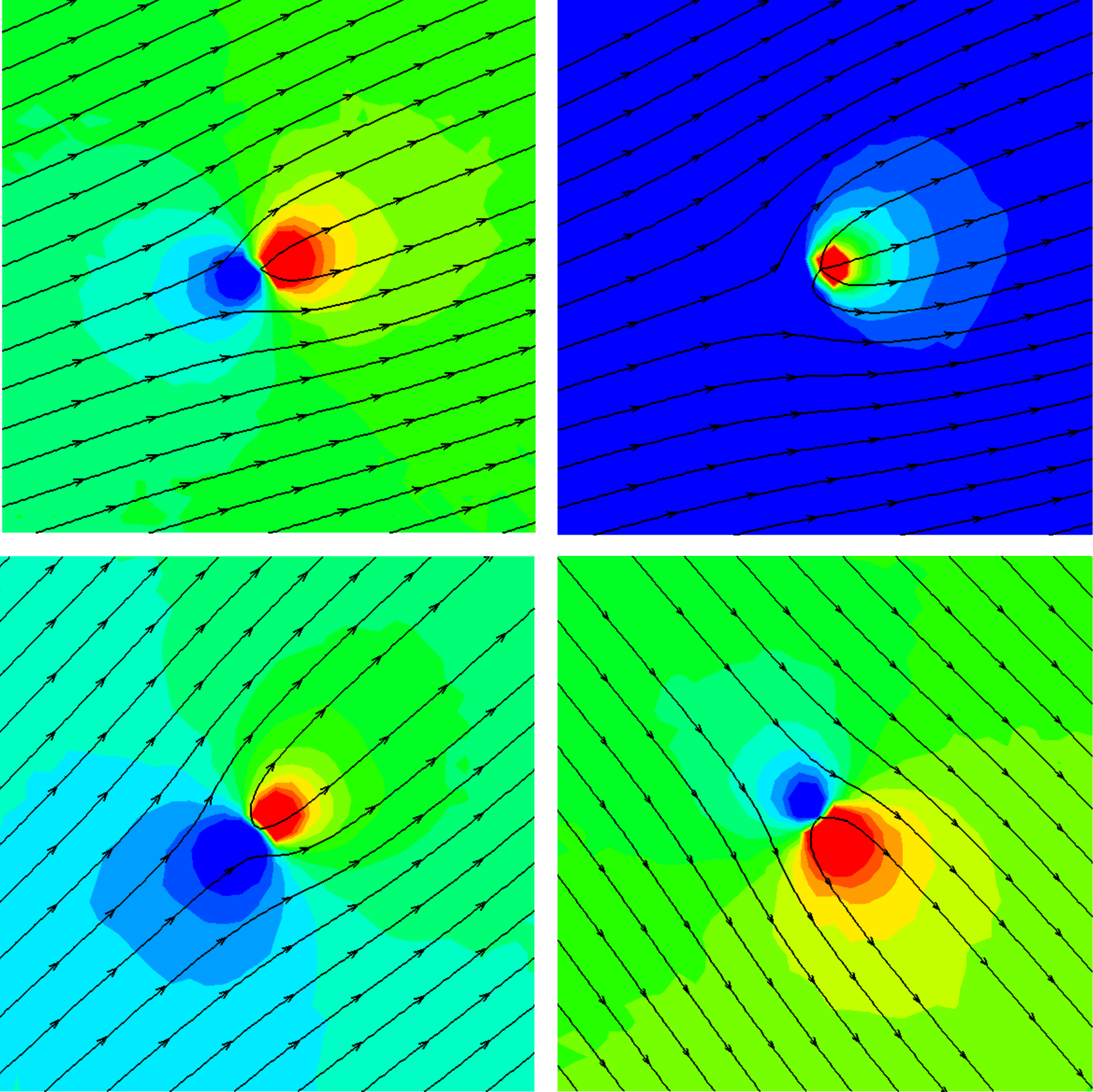}
        \caption{Constrained GANs, epoch 120}
    \end{subfigure}
    \caption{
    Example velocity fields (streamlines and color contours) generated standard GANs (left column: panels a, c, e)and constrained GANs (right column: panels b, d, f). Samples at different training epochs (40, 80, and 120) are shown to illustrate the gradual improvement of generated samples with increased training epoch.
    } 
    \label{fig:generated-v}
\end{figure}

The generated velocity fields by the standard GAN and constrained GAN are presented in Fig.~\ref{fig:generated-v} at different epochs of training. A learning rate of $\eta = 0.005$ and a tolerance value of $\varepsilon = 0$ are used. It can be observed that the quality of the velocity fields improves as the training proceeds for both the standard GAN and the constrained GAN. This is evident from the improved smoothness of the velocity magnitude contours downstream of the source. From the visual inspection, one can hardly see any qualitative differences among the samples generated by the standard GAN and the constrained GAN. It is encouraging to see that both seem to generate ``realistic'' potential flow velocity fields that are qualitatively similar to the training samples (Fig~\ref{fig:N-training}a). It is worth emphasizing again that the training data consists of flow fields (velocity and pressure) on a discretized mesh of $32 \times 32$ grid points, which apparently reside in a high dimensional space ($32 \times 32 \times 2$, where the number 2 corresponds to the two velocity components). Recall the fact that all fields are controlled by three parameters and are generated through the potential in Eq.~\eqref{eq:phi} is transparent to the GAN, which only sees the mesh-based field instead. From this perspective, it is encouraging that the GANs (either standard or constrained) learned the correct distribution of the data (i.e., a superposition of a uniform flow and a source flow), at least approximately. This observation supports the theoretical statement that the generator is able to replicate all statistics in the training data if global optimum is achieved in the training~\cite{goodfellow2014generative}.

\begin{figure}
\centering
    \includegraphics[width=0.6\textwidth]{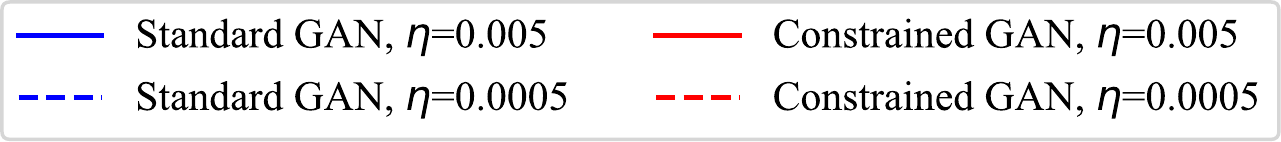}
    
    \centering
    \begin{subfigure}[b]{0.48\textwidth}
    \centering
        \includegraphics[width=1\textwidth]{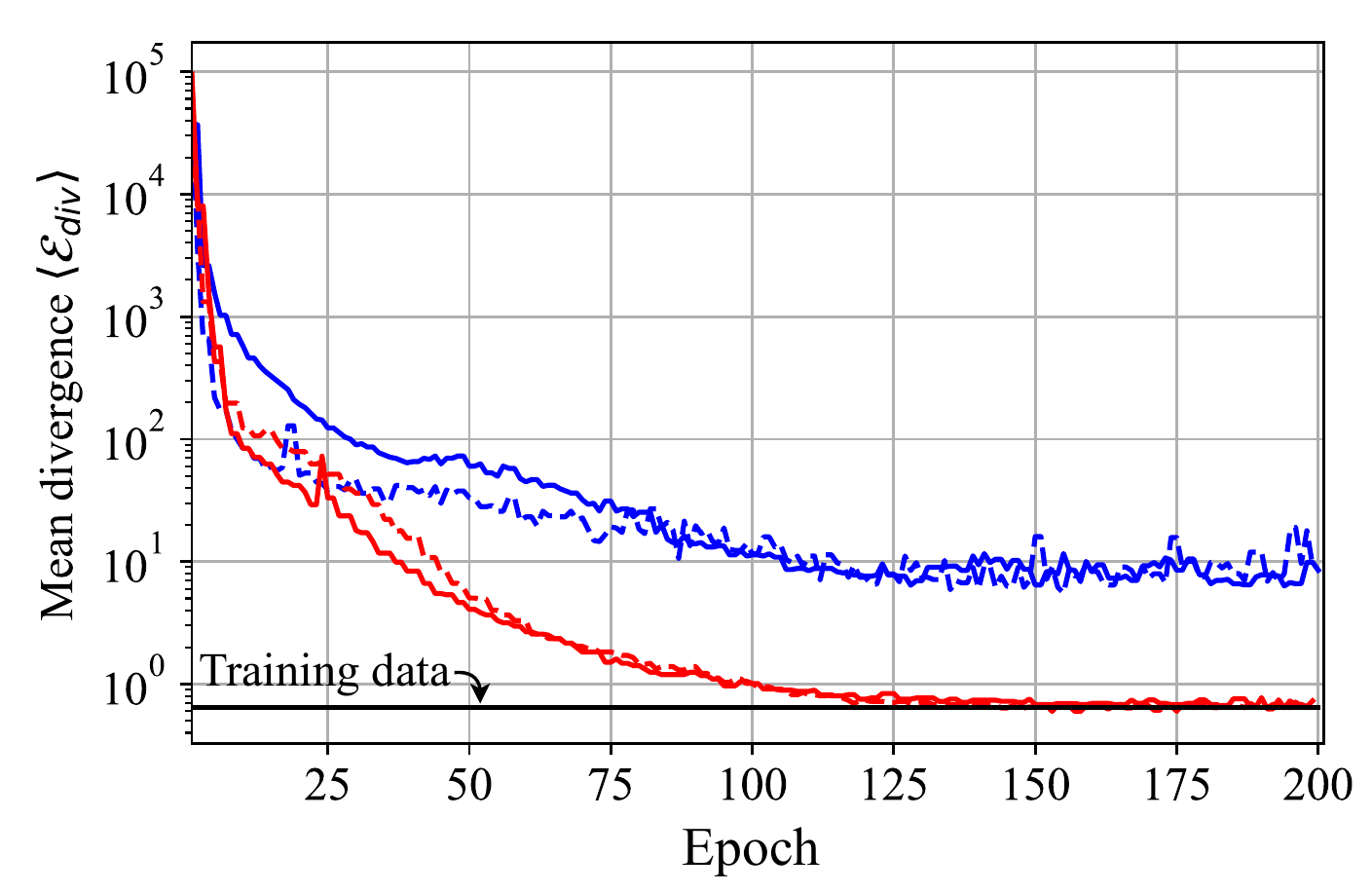}
        \caption{Mean divergence $\langle \mathcal{E}_\text{div}\rangle$}
        \label{fig:flow-div-smooth-1}
    \end{subfigure}
    \begin{subfigure}[b]{0.48\textwidth}
    \centering
        \includegraphics[width=1\textwidth]{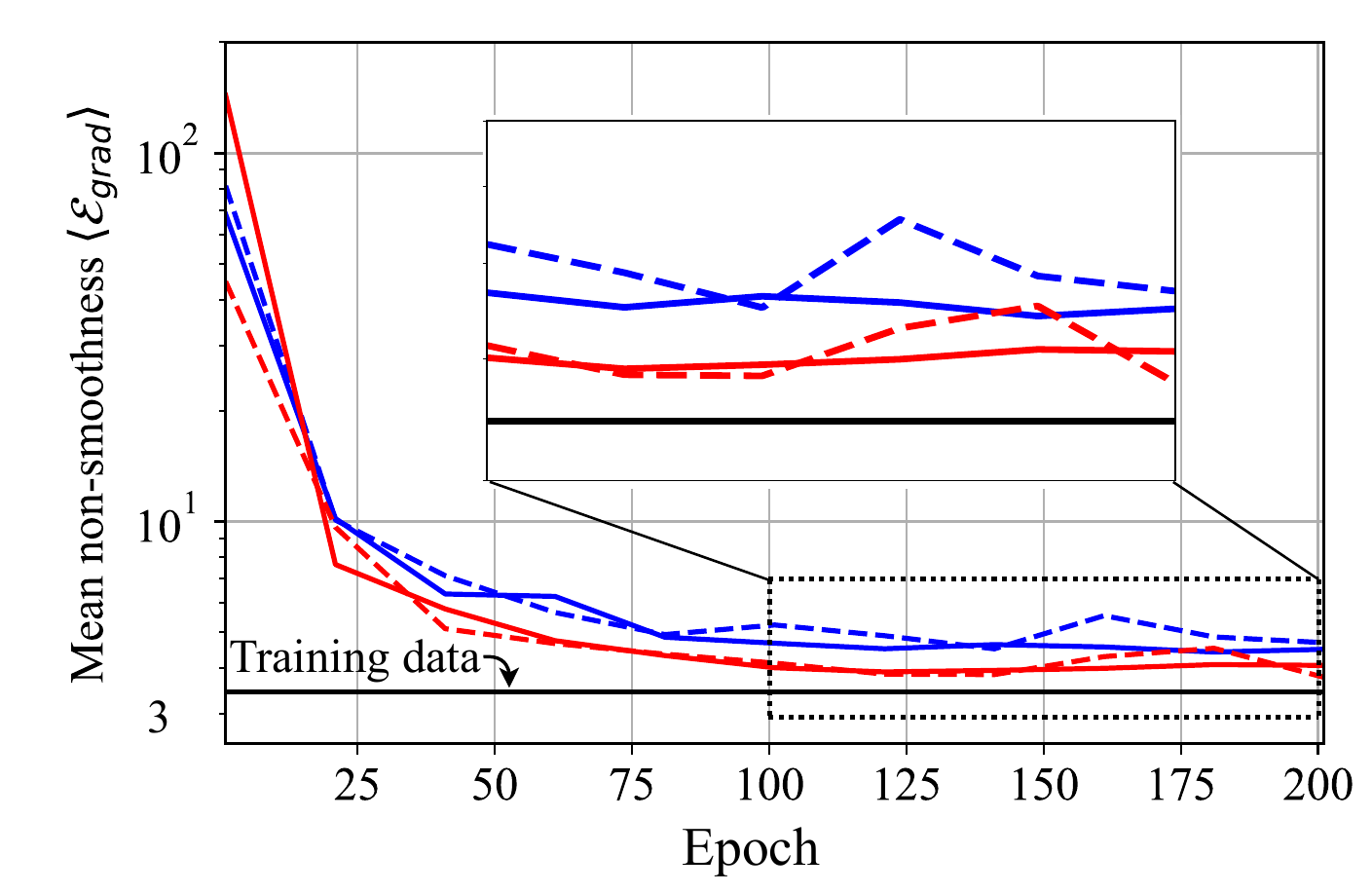}
        \caption{Mean non-smoothness $\langle \mathcal{E}_\text{grad} \rangle$}
        \label{fig:flow-div-smooth-2}
    \end{subfigure}
    \caption{
    Quantitative assessment of velocity field samples generated by the standard and constrained GANs at different learning rates $\eta = 0.005$ and $0.0005$. Two metrics are shown: (a) mean divergence $\langle \mathcal{E}_\text{div} \rangle$ of the velocity fields as defined in Eq.~\eqref{eq:ediv}, and (b) the mean non-smoothness $\langle \mathcal{E}_\text{grad} \rangle$ as defined in Eq.~\eqref{eq:egrad}. The corresponding divergence and non-smoothness of the training velocity fields are indicated for comparison (in horizontal black solid lines).
    }
    \label{fig:div-trainingprocess}
\end{figure}

The generated velocity fields are further analyzed by computing the means of the velocity divergence $\langle \mathcal{E}_\text{div} \rangle$ and the non-smoothness $\langle \mathcal{E}_\text{grad} \rangle$ in order to assess the generated samples more quantitatively. The results are presented in Fig.~\ref{fig:div-trainingprocess}. The quantitative assessment suggests that the constrained GAN outperforms the standard GAN. For both standard GAN and constrained GAN, the velocity divergence decreases as the training proceeds (i.e., as the epoch number increases). However, even after convergence (at about 100 epochs) the standard GAN sample still have divergence almost one order of magnitude larger than that of the training samples (see Fig.~\ref{fig:flow-div-smooth-1}).
In contrast, the constrained GAN reaches the same levels of divergence as that of the training samples. This is expected since the divergence-free constraint is embedded in the generator loss function. Nevertheless, an unexpected incidental consequence is that the constrained GAN also generated smoother velocity fields than the standard GAN, although still not as smooth as the training velocity fields (see Fig.~\ref{fig:flow-div-smooth-2}).
Finally, in these tests we use two learning rates, $\eta =0.005$ and $0.0005$ and the results are found to be similar with both settings after convergence. Therefore, the observations above are insensitive to choice of the algorithmic parameter. However, the well-tuned learning rate ($\eta =0.005$) does lead to faster convergence for both the standard GAN and the constrained GAN. 

\begin{figure}
\centering
     \begin{subfigure}[b]{0.8\textwidth}
    \centering
        \includegraphics[width=1\textwidth]{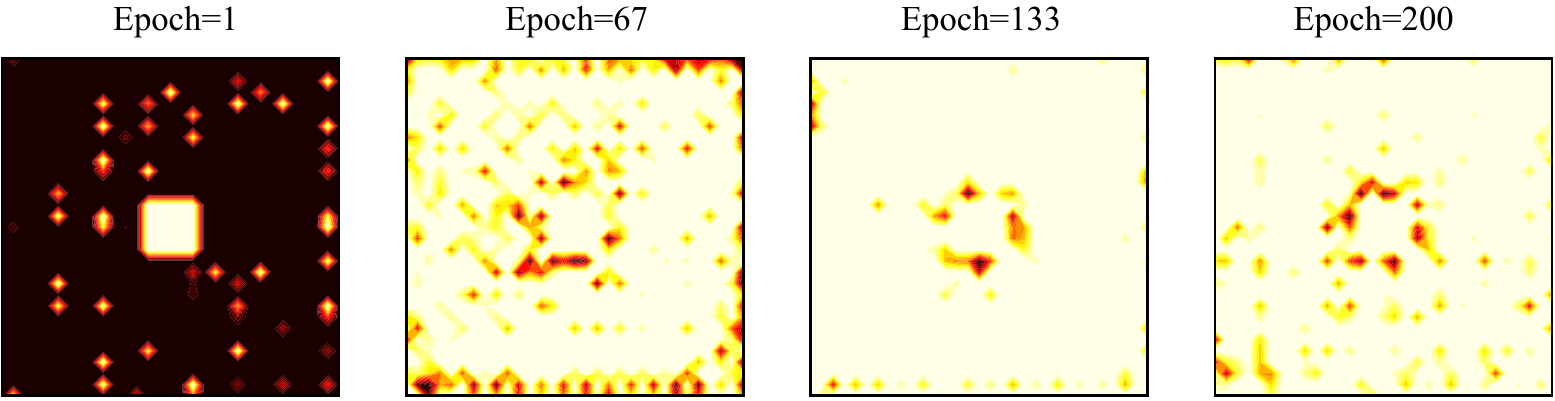}
        \caption{Standard GAN}
    \end{subfigure}
    \begin{subfigure}[b]{0.8\textwidth}
    \centering
        \includegraphics[width=1\textwidth]{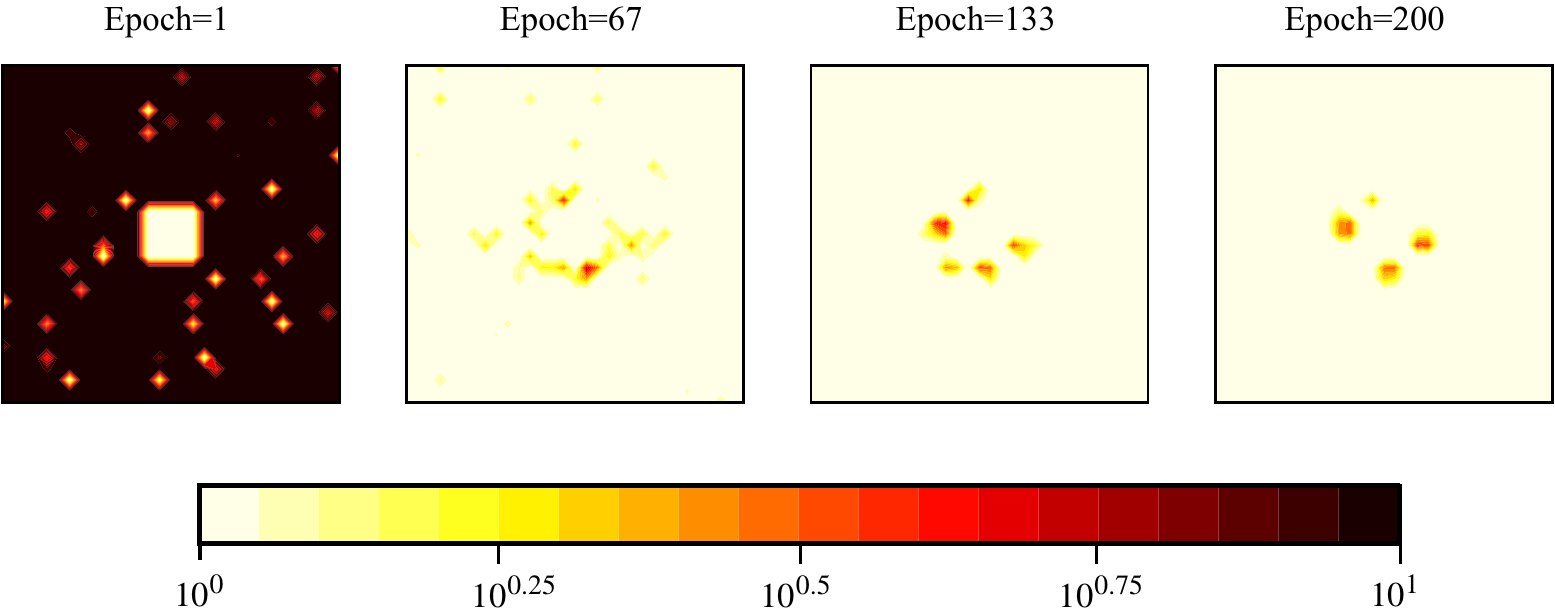}
        \caption{Constrained GAN}
    \end{subfigure}
    \caption{
    The contours of logarithmic divergence magnitude at different epochs of training for (a) standard GAN and (b) constrained GAN.
    The divergence near the source location $z_0=(0, 0)$ at the center of the domain is set as zero to avoid singularity. The optimal learning rate $\eta = 0.005$ is used.
    }
    \label{fig:divergence-fields}
\end{figure}

Furthermore, the divergence fields (depicted as color contours) at different stages of the training (epochs 67, 133, and 200) are compared for the standard and the constrained GANs in Fig.~\ref{fig:divergence-fields}. It can be clearly seen from these contours that the ``hotspots'' of large divergences (i.e., mass conservation error, darker does) decrease more rapidly for the constrained GAN with increasing epoch. This observation corroborates the time history of total divergence shown in Fig.~\ref{fig:div-trainingprocess}a above. That is, the constrained GAN is better at reducing the divergence in the generated velocity fields, which is evidently due to the explicit constraints applied to penalize large divergence values.

In summary, the results shown in Figs.~\ref{fig:generated-v}--\ref{fig:divergence-fields} suggest that both the standard and constrained GANs can capture the overall distribution of training data, and that the training convergences are robust to the tuning parameter (learning rate $\eta$) for both of GANs. However, the constrained GAN clearly outperforms standard GAN in the following aspects:
\begin{enumerate}[(a)]
    \item the constrained GAN conforms better to the physical constraint (thanks to the embedded constraint in generator loss function), 
    \item the constrained GAN generates velocity fields of comparable smoothness as the training data, which are smoother than samples generated by the standard GAN, , even though the smoothness is not explicitly enforced in the loss function, and 
    \item the constrained GAN has faster convergence than the standard GAN.
\end{enumerate}

\begin{figure}
    \centering
    \includegraphics[width=0.5\textwidth]{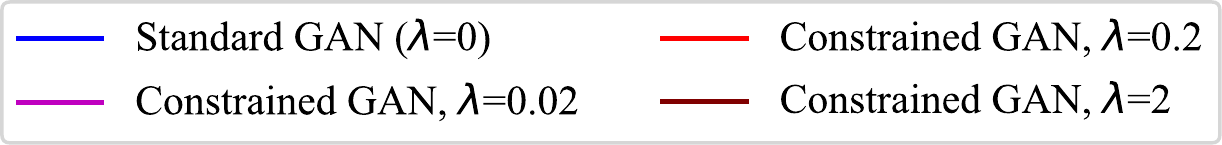}
    \includegraphics[width=0.5\textwidth]{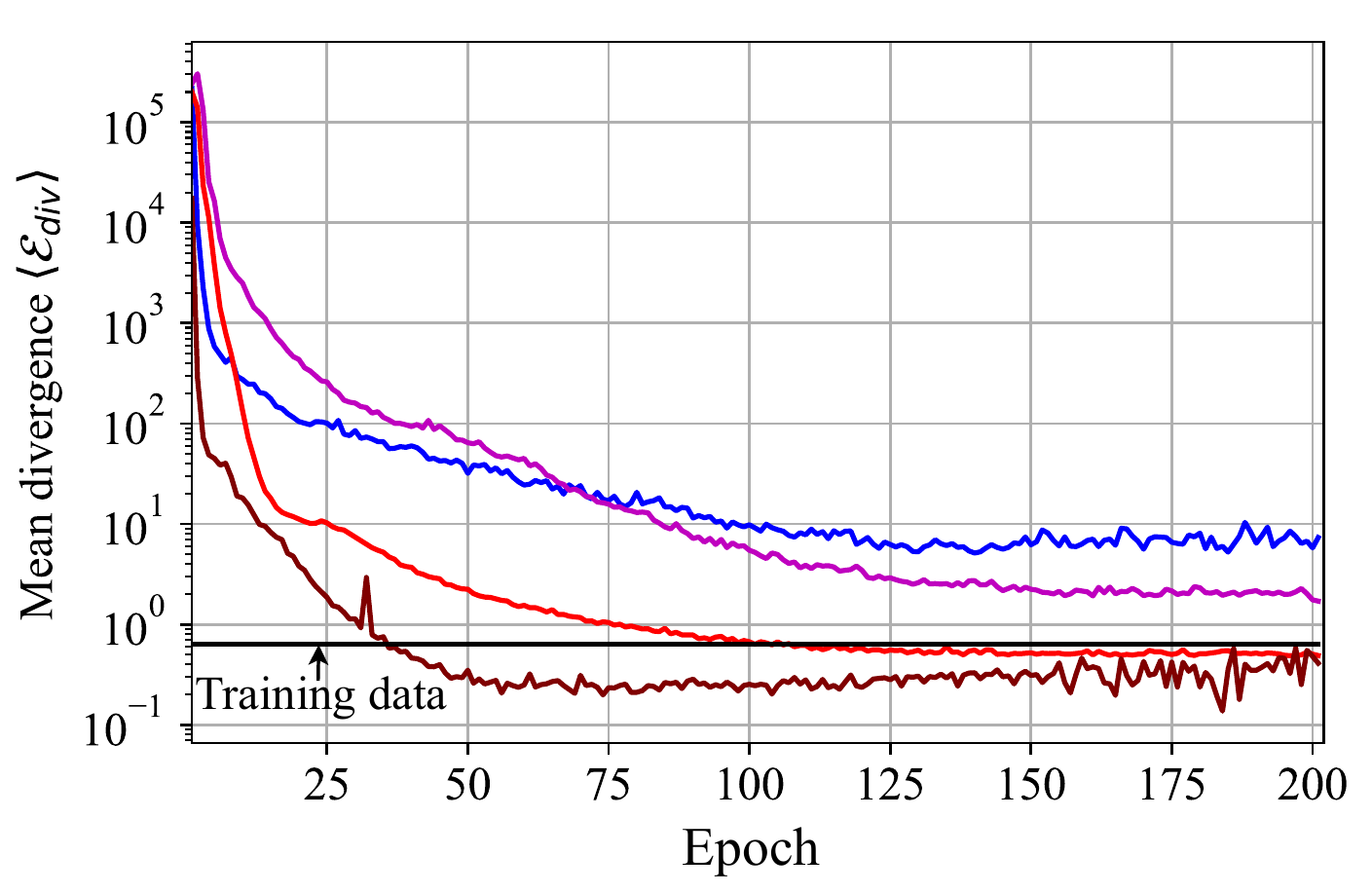}
    \caption{The mean of divergence of generated velocity fields by using constrained GANs with different weights $\lambda$ in constraint term during the training process. Training history of the mean divergence for constrained GAN with three weights $\lambda = 0.02$, $0.2$ and $2$ and that for standard GAN (corresponding to $\lambda = 0$) are shown.}
    \label{fig:div-differentweight}
\end{figure}

The physics-constrained GAN as proposed in this work aims to satisfy two objectives simultaneously with the two term in the loss function in Eq.~\eqref{eq:loss-PIGANs}, i.e., to generate sample that (i) conform to the training data distribution and (ii) satisfy the physical constraint. The relative priority is determined by the relative weight of the terms as controlled by  parameter~$\lambda$.  
A suitable parameter~$\lambda$ must be chosen to satisfy both objectives as outlined above. Otherwise, one term may completely dominate the other, and the generated sample would either fail to conform to the data distribution or violate the physical constraints. We found a good rule of thumb is to choose $\lambda$ so that the two terms in the loss function (standard GAN loss and physical constraint loss) are of comparable magnitude, which should be dynamically monitored. We have tested three different choices of weight parameters, $\lambda = 0.02$, $0.2$, and $2$. The mean divergence at different epochs of training is presented in Fig.~\ref{fig:div-differentweight}. The generated samples have a divergence comparable to the training data with the optimal weight $\lambda = 0.2$.  It can be seen that the divergence-free constraint is still somewhat satisfied with a smaller weight $\lambda = 0.02$. In this case the loss function is dominated by the standard GAN loss and the physical constraint is not effectively enforced. As expected, samples generated by the standard GAN (corresponding to $\lambda = 0$) depart from the divergence-free condition even further.  On the other hand, with a large weight of $\lambda = 2$, the divergence is an order of magnitude smaller even than the training sample.  However, generated velocity fields  (not shown here) look dramatically different from a potential velocity field, which is clearly undesirable.  Therefore, it is essential to choose the weight parameter $\lambda$ such that the two terms in the loss function are balanced. In this way, the generated samples will both conform to the data distribution and satisfy the physical constraint.

\begin{figure}
    \centering
    \includegraphics[width=0.6\textwidth]{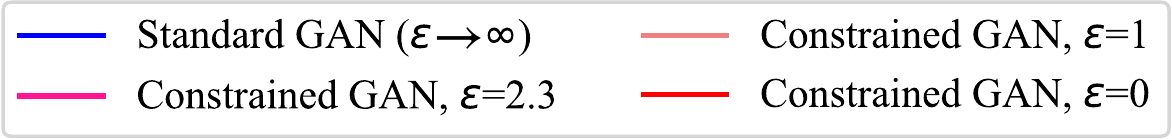}
        \includegraphics[width=0.6\textwidth]{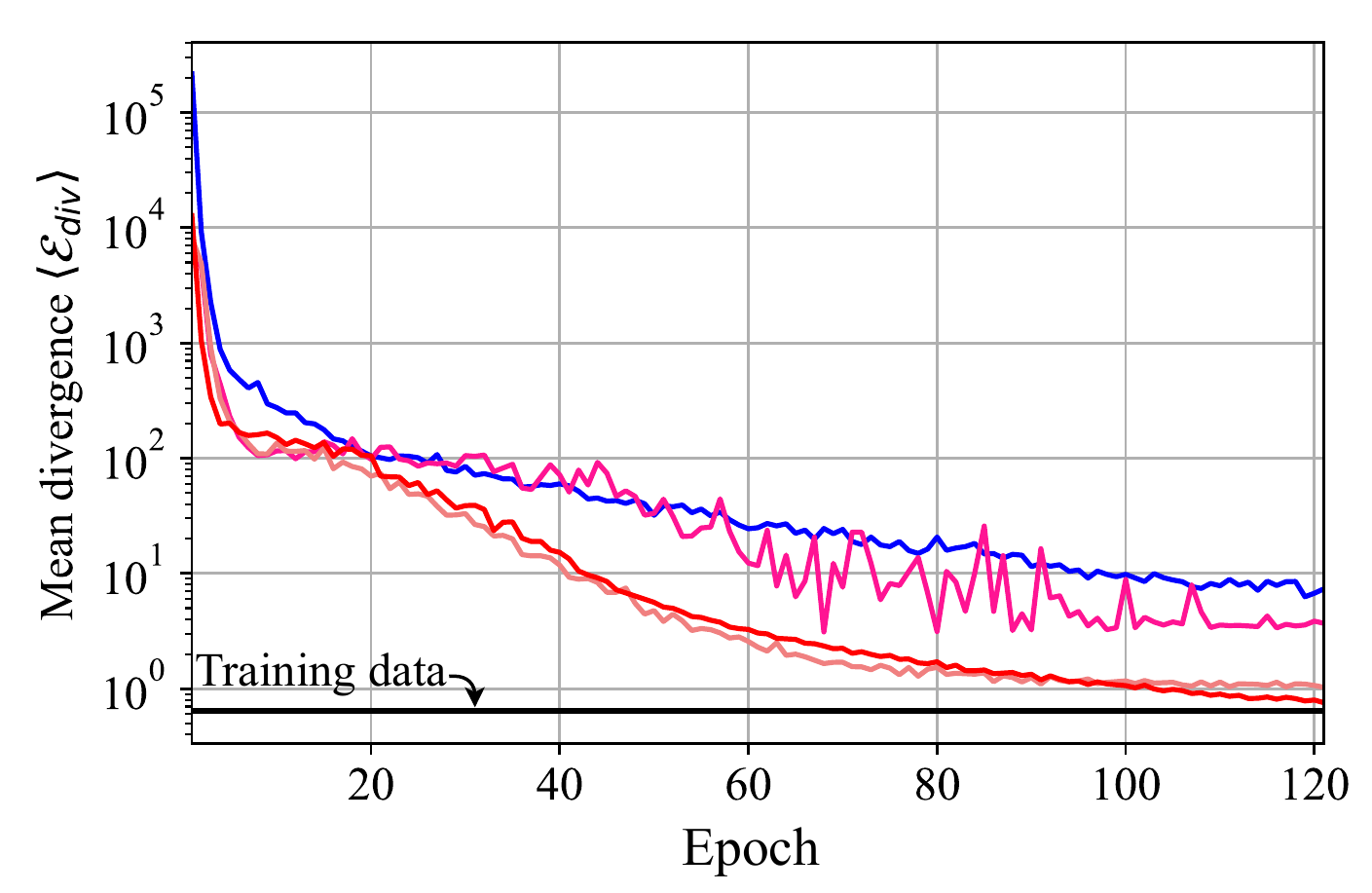}
    \caption{The mean of divergence value of generated velocity fields in the process of training of constrained GANs with level of imprecise constraints. Different values of $\varepsilon$ ($0$, $1$, and $2.3$) correspond to different levels of imprecise constraints. The larger $\varepsilon$ is, the less accurate the constraint is enforced.}
    \label{fig:div-differentsoft}
\end{figure}

A similar parametric study is performed on the constraint tolerance $\varepsilon$ with three values ($\varepsilon =0$, $1$, and $2.3$) investigated. The mean divergence of the velocities corresponding to these parameter choices are presented in Fig.~\ref{fig:div-differentsoft}. In this case the standard GAN corresponds to an infinitely large tolerance $\varepsilon \to \infty$. Both $\varepsilon = 0$ and $\varepsilon = 1$ lead to very well satisfied divergence-free condition.
At $\varepsilon = 2.3$, the divergence-free condition is not as well satisfied, while the standard GANs generate fields with even larger divergence.
This is similar to what was observed for the weight parameter, where a larger weight parameter $\lambda$ places more emphasis on the physical constraint, except that in this case it is not possible to dominate the standard GAN loss term with the smallest tolerance $\varepsilon = 0$. Overall, it can be concluded from Fig.~\ref{fig:div-differentsoft} that the imprecise constraint can improve the training of GANs and make the generated samples conform to the specified physical constraint.

Significant fluctuations are observed in Fig.~\ref{fig:div-differentsoft} in the time history of divergence for the case of $\varepsilon=2.3$. This indicates that the training process is unstable. This can be explained by the discontinuous nature of the loss function Eq.~\eqref{eq:loss-div}: it is zero if the divergence magnitude $|\nabla \cdot \bm{v}| < \varepsilon$ and otherwise it is $|\nabla \cdot \bm{v}|$. In other words, it is active only when the divergence exceeds $\varepsilon$. 
Given that the mean divergence of the training velocity fields is 0.64 (indicated as horizontal black line in Fig.~\ref{fig:div-differentsoft}) due to the numerical discretization error, the tolerance value $\varepsilon = 2.3$ is a rather soft constraint. Consequently, the loss function is active very intermittently, which causes discontinuous changes of the loss function and its derivatives and thus it leads to instability. In contrast, in other cases, e.g., $\varepsilon=0$ (exact constraint), $\varepsilon = 1$ (imprecise constraint), or $\varepsilon \to \infty$ (the standard GAN), the constraint is either almost always active or
almost always inactive, and thus they are free from such intermittency-induced instabilities. It can be seen that the regularization terms in loss functions can influence the training of neural networks in unexpected ways, and further research on enforcing constraints in GANs is warranted.

\section{Conclusion}
\label{sec:conclusion}

In this work, we proposed a general approach of embedding constraints in GANs for emulating physical systems. 
The physical constraints are embedded to the loss function of the generator to help the weaker party in the two-player game, which help the training to achieve equilibrium faster.
This is motivated by the observation that the constraints reduce the effective dimension of the search space for the weight optimization and thus accelerate the training convergence. Moreover, we extend the proposed framework of constrained GANs to incorporate imprecise constraints. This is motivated by the fact that many physical constraints are not known exactly in practical applications. Two simple yet representative test cases are used to demonstrate the merits of physics-constrained GANs: (i) generating circle on a plane, which features geometrical constraints and (ii) generating potential flow velocity fields, which have differential constraints. We show that enforcing constraints in the generator both accelerates the convergence of GANs training and improves the quality of generated samples. The proposed method is generally applicable to different physical systems and will facilitate the application of using GANs to emulate complex physical systems. A limitation of this work is that it does not account for the cases where the training data themselves do not conform to the constraints (e.g., due to pollution or measurement noises). Such scenarios are of equal importance and thus are worth further investigations.

\section*{Acknowledgment}
The computational resources used for this project were provided by the Advanced Research Computing (ARC) of Virginia Tech, which is gratefully acknowledged.

\end{document}